\documentstyle[preprint,epsfig,aps]{revtex}
\begin{document}
\draft
\preprint{\begin{tabular}{l}
\hbox to\hsize{October, 1998 \hfill KAIST-TH 98/15}\\[-3mm]
\hbox to\hsize{hep-ph/9810336  \hfill KIAS-P98021}\\[-3mm]
\hbox to\hsize{\hfill SNUTP 98-078 }\\[-3mm] 
\end{tabular} }

\bigskip

\title{Possible new physics signals \\ in $b\rightarrow s \gamma$ 
and $b\rightarrow s l^+ l^-$}
\author{Yeong Gyun Kim and P. Ko }
\address{Department of Physics, 
KAIST \\ Taejon 305-701, Korea}
\author{ Jae Sik Lee }
\address{School of Physics, KIAS 
\\ Seoul 130-012, Korea}
\maketitle
\begin{abstract}
We consider possible new physics contributions to $b\rightarrow s l^+ l^-$ 
assuming the new physics modifies (chromo)magnetic and electric form factors 
in  $b\rightarrow s \gamma^*$ and $b\rightarrow s  g$ with the same chirality 
structure as in the standard model.
Parametrizing the new physics effects on $b\rightarrow s \gamma^*$ and 
$b\rightarrow s  g$ in terms of four real parameters, one finds that 
there are enough region of parameter space in which the measured branching 
ratio for $B \rightarrow X_s \gamma$ can be accomodated, and the predicted 
CP violation effect could be as large as $\sim 30 \%$. Moreover, the 
branching ratio and the forward-backward asymmetry of a lepton  in $B 
\rightarrow X_s l^+ l^-$ and the tau polarization asymmetry in 
$B \rightarrow X_s \tau^+ \tau^-$ can be deviated from the SM predictions 
by a factor of $\sim 2$, which can be accessible at B factories. 
We also discuss these observables in a specific class of supersymmetric 
models with gluino-mediated flavor changing neutral current (FCNC).  
\end{abstract}


\newpage
\narrowtext
\tighten

\section{Introduction}

The missions of $B-$factories under constructions are (i) to test the CP 
violation in the Standard Model (SM) {\it \'{a} la} Kobayashi-Maskawa scheme 
\cite{km}, and (ii) to find out any new flavor violation and especially new 
source of  CP violation beyond the KM phase in the SM with three generations. 
The latter is well motivated by the fact that the KM phase in the SM may not 
be enough to generate the baryon number asymmetry in the universe. 
In terms of physics view point, the second mission seems more exciting one, 
since it could uncover a veil beyond the SM and provide an ingredient that is 
necessary to explain baryon number asymmetry of the universe. 
Then, one has to seek for a possible signal of new physics in rare decays of 
$B-$mesons and CP violation therein. One could choose his/her own favorite 
models to work out the consequences of such model to the physics issues that
could be investigated at B factories. Or one could work in the effective field 
theory framework, in a manner as much as model-independent as possible. 
In the following, we choose the second avenue to study the possible signals
of new physics that could be studied in detail at $B$ factories. Then we give 
explicit examples (that satisfy our assumptions made in the model independent 
analysis) in supersymmetric (SUSY) models with gluino-mediated $b\rightarrow 
s$ transition.  

If one considers the SM as an effective field theory (EFT) of more fundamental
theories below the scale $\Lambda$, the new physics effects will manifest 
themselves in higher dimensional operators (dim $[O] \geq 5$) that are 
invariant under the SM gauge group.  Several groups have made a list of 
dimension-5 and dimension-6 operators in the last decade \cite{buch}. 
Assuming the lepton and baryon number conservations, there are about 
80 operators that are independent with each other. It would be formidable 
to consider all of such operators at once, even if we are interested in 
their effects in $B$ physics. However, if we restrict to 
$b\rightarrow s \gamma$, only two operators become relevant :
\begin{equation}
a_L {v \over \Lambda^2}~ \bar{s}_L \sigma_{\mu\nu} b_R~F^{\mu\nu},~~~
{\rm and}~~~
a_R {v \over \Lambda^2} \bar{s}_R \sigma_{\mu\nu} b_L~F^{\mu\nu},
\end{equation}
after the electroweak (EW) symmetry breaking ($v$ is the Higgs vacuum 
expectation value). Here $a_{L,R}$'s are dimensionless coefficients.  
Thus the above operators can be recasted into the following form 
\footnote{We follow the convention of Ref.~ \cite{buras}} :
\begin{equation}
H_{\rm eff} (b \rightarrow s \gamma) = - { G_F \lambda_t \over 
\sqrt{2}} \left[ C_{7L} O_{7L} + C_{7R} O_{7R} \right], 
\end{equation}
where $\lambda_t = V_{ts}^* V_{tb} (= - A \lambda^2$ in the Wolfenstein 
parametrization \cite{wolfenstein})  and  
\begin{equation}
O_{7L}  = {e \over 4 \pi^2 }~m_b \bar{s}_L^{\alpha} 
\sigma^{\mu\nu} b_R^{\alpha}~F_{\mu\nu}.
\end{equation}
The operator $O_{7R}$ is obtained from $O_{7L}$ by the exchange 
$( L \leftrightarrow R)$.  Similarly one can expect a new physics 
contribution to $b\rightarrow s g$ : 
\begin{equation}
H_{\rm eff} (b \rightarrow s  g) = - { G_F \lambda_t \over 
\sqrt{2}} \left[ C_{8L} O_{8L} + C_{8R} O_{8R} \right],
\end{equation}
where 
\begin{equation}
O_{8L}  = {g_s \over 4 \pi^2 }~m_b \bar{s}_L^{\alpha} 
\sigma^{\mu\nu} T^a_{\alpha \beta} b_R^{\beta}~G^a_{\mu\nu},
\end{equation}
and $O_{8R}$ is obtained from $O_{8L}$ by the exchange $(L\leftrightarrow R)$.
These two processes $b\rightarrow s\gamma$ and $b\rightarrow s g$ are unique 
in the sense that they are described in terms of only two independent operators
$O_{7(8)L}$ and $O_{7(8)R}$ whatever new physics there are. This fact makes it
easy to study these decays in a model indepdent manner \cite{kn98}. 

The SM predictions for the $C_{7,8}$ at the $M_W$ scale are 
(in the limit $m_s = 0$)
\begin{eqnarray}
C_{7L}^{\rm SM} (M_W) & \approx & -0.22, 
\nonumber  
\\
C_{7R}^{\rm SM} (M_W) & = & 0,
\nonumber  
\\
C_{8L}^{\rm SM} (M_W) & \approx & -0.12, 
\nonumber  
\\
C_{8R}^{\rm SM} (M_W) & = & 0.
\end{eqnarray}
Note that $C_{7(8)R}^{\rm SM}$ in the SM is suppressed compared to 
$C_{7(8)L}^{\rm SM}$ by $m_s/m_b$, because $W$ boson couples only to the 
left-handed fermions. Such terms proportional to $m_s$ will be  neglected in 
our work by setting $m_s = 0$ whenever they appear.  On the 
other hand, this chirality suppression needs not be true in  the presence of 
new physics such as Left-Right symmetric (LR) model or in a certain 
class of supersymmetric models with specific flavor symmetries. 
Such new physics contributions can be parametrized in terms of four 
complex parameters,
\begin{eqnarray}
C_{7L}^{\rm New} (m_W) & = & C_{7L}^{\rm SM} (m_W) ~( \xi_7 - 1), 
\nonumber  
\\
C_{8L}^{\rm New} (m_W) & = & C_{8L}^{\rm SM} (m_W) ~( \xi_8 - 1),
\nonumber  
\\
C_{7R}^{\rm New} (m_W) & = & C_{7L}^{\rm SM} (m_W) ~\xi_7^R , 
\nonumber  
\\
C_{8R}^{\rm New} (m_W) & = & C_{8L}^{\rm SM} (m_W) ~\xi_8^R ,
\end{eqnarray}
where $\xi_{7,8}^{(R)}$ are new complex numbers, whose  phases parametrize 
the effects of the new sources of CP violation beyond the KM phase in the 
SM. The SM case corresponds to $\xi_{7,8} = 1$ and $\xi_{7,8}^R = 0$.       
It is  convenient to define the ratio $\chi$ as following :
\begin{equation}
\chi \equiv (\xi_8 - 1)/(\xi_7 - 1)=
\frac{C_{8L}^{\rm New}(M_W)/C_{8L}^{\rm SM}(M_W)}{C_{7L}^{\rm
New}(M_W)/C_{7L}^{\rm SM}(M_W)}.
\end{equation}
In many interesting cases, this parameter $\chi$ is real \cite{kn98} 
as assumed in this work. 

Implications of new physics contributions to $b\rightarrow s g$ have been
discussed by various group in conjunction with the possible solutions for the 
discrepancies between theoretical expectations and the data on  the 
semileptonic branching ratio of and the missing charms in $B$ meson decays,  
and the unexpectedly large branching ratio for $B \rightarrow \eta^{'} + X_s$.
It has been advocated that $B(B\rightarrow X_{sg}) 
\approx 10 \% [ \sim 50 \times B_{\rm SM} (B \rightarrow X_{sg} )]$ 
can solve these problems simultaneously \cite{bsg}. 
However,  this claim is now being  challenged by the new 
measurement  $B( B \rightarrow X_{sg} ) <  6.8 \% ~(90 \%$ CL) \cite{coan}.  
In this work, we impose this new experimental data, rather than assume that 
the $B(B\rightarrow X_{sg} )$ is large enough to solve the aformentioned 
puzzles in $B$ decays.

In the presence of new physics contributions to $b\rightarrow s \gamma$, 
there should be also generic new physics contributions to $b\rightarrow s l^+ 
l^-$ through electromagnetic penguin diagrams. This effect will modify the 
Wilson coefficient $C_9$ of the dim-6 local operator $O_9$ :
\begin{equation}
H_{\rm eff} (b\rightarrow s ll) \supset 
H_{\rm eff} (b\rightarrow s \gamma ) 
- { G_F \lambda_t \over \sqrt{2}} ~\left[ C_9 O_9 + C_{10} O_{10} \right] ,
\end{equation}   
where 
\begin{equation}
O_9 = {e^2 \over 4 \pi^2} \left( \overline{s}_L \gamma_{\mu} b_L \right)~
( \overline{l} \gamma^{\mu} l ),
~~~
O_{10} = {e^2 \over 4 \pi^2} \left( \overline{s}_L \gamma_{\mu} b_L \right)~
( \overline{l} \gamma^{\mu} \gamma_5 l ).
\end{equation}
In the SM, the Wilson coefficients $C_{9,10}$'s are given by
\begin{equation}
C_9^{\rm SM} ( M_W ) \approx 2.01,
~~~
C_{10}^{\rm SM} ( M_W ) \approx 4.55.
\end{equation}
Let us parametrize the new physics contribution to $C_{9}$  
in terms of $\xi_9$ (or $\chi^{'}$) as following :
\begin{eqnarray}
C_9^{\rm New} (M_W)&=& C_9^{\rm SM} (M_W)(\xi_9 - 1) = C_9^{\rm SM} (M_W)\chi^{'} 
( \xi_7 - 1 ),\nonumber \\
\chi^{'}&=&
\frac{C_{9}^{\rm New}(M_W)/C_{9}^{\rm SM}(M_W)}{C_{7L}^{\rm
New}(M_W)/C_{7L}^{\rm SM}(M_W)}.
\end{eqnarray}
Since we assume that the new physics modifies only $b\rightarrow s\gamma^*$ 
and $b\rightarrow s g$, we have $C_{10} (M_W) = C_{10}^{\rm SM} (M_W)$  
\footnote{The $Z$ penguin contribution to $b\rightarrow s l^+ l^-$ is 
supressed relative to the photonic penguin by a factor of $O(M_{ll}^2/M_Z^2)$, 
and thus neglected in this work.}.
There is no model-independent relation between $\xi_7$ and $\xi_9$, although
they are generate by the same Feynman diagrams for $b \rightarrow s \gamma^*$.
In Sec. IV, we will encounter examples for both $\chi^{'} = 0$ and $\chi^{'}
\neq 0$ in general SUSY models with gluino-mediated flavor changing neutral 
current (FCNC).  
In principle, there are many more dim-6 local operators that might contribute 
to $b\rightarrow s l^+ l^-$ \cite{cskim}.  
In the presence of so many new parameters, it 
is  difficult to figure out which operators are induced by new physics,
since we are afforded only a few physical observables, such as $B ( B
\rightarrow X_s \gamma), B( B \rightarrow X_s l^+ l^-), A_{\rm FB} (B
\rightarrow X_s l^+ l^-)$ and the tau polarization asymmetry $P_{\tau}$ in 
$B \rightarrow X_s \tau^+ \tau^-$. Therefore, it would be
more meaningful to consider the simpler case before we take into account the
most general case to figure out which operators are significantly affected by
new physics.

Up to now, we considered $O_{7,8(L,R)}, O_{9}$ and $O_{10}$ relevant to 
$B \rightarrow X_s \gamma, X_{sg}, X_s l^+ l^-$, assuming 
new physics significantly contributes
to $b\rightarrow s \gamma^*$ 
and $b\rightarrow s  g$ through dim-5 operators, Eqs.~(2)--(5). In doing 
so, five more complex numbers ($\xi_{7,8(L,R)},\xi_9$) have been introduced.   
If we further assume that the new physics does not induce new operators that 
are absent in the SM, we can drop $O_{(7,8)R}$ by setting $\xi_{7,R} = 
\xi_{8,R} = 0$, thereby reducing the number of new parameters characterizing 
new physics effects into three complex numbers $\xi_{7,8,9}$'s (or, 
equivalently $\xi_7, \chi$ and $\chi^{'})$. Still the number of new 
parameters  are larger than the physical observables at our disposal. However,
in many interesting cases (and especially SUSY models with gluino-mediated 
$b\rightarrow s$ transition that is to be described in Sec. IV), it turns out 
that both $\chi$ and $\chi^{'}$ are real. Therefore, we will assume that both 
$\chi$ and $\chi^{'}$ are real hereafter, and we are end up with 4 real 
parameters, which we choose to be $| \xi_7 |, {\rm Im}~ (\xi_7), \chi$ and 
$\chi^{'}$. Then we can overconstrain these parameters from the following 
observables :
\begin{itemize}
\item the branching ratio for $B\rightarrow X_s \gamma$ relative to the SM 
prediction $(R_{\gamma})$
\item the direct CP violation in $B\rightarrow X_s \gamma$ ($A_{\rm CP}^{b
\rightarrow s \gamma} \equiv A_{\rm CP})$
\item the branching ratio for $B \rightarrow X_{s g}$ relative to the SM 
prediction $(R_{g})$
\item the branching ratio for $B\rightarrow X_s l^+ l^-$ relative to the SM
prediction $( R_{ll})$
\item the forward-backward asymmetry in $B\rightarrow X_s l^+ l^-$ 
($ A_{FB}^{b\rightarrow sll}$)
\item the tau polarization asymmetry  in  $B\rightarrow X_s \tau^+ \tau^-$ 
($ P_{\tau}$) 
\end{itemize}

\noindent
At this point, it is timely to recall that there have been several works on 
the model-indepedent determination of the Wilson coefficients, $C_{7,(8),
9,10}$ from $R_{\gamma}$ and the kinematic distributions in $R_{ll} $
\cite{ali} -- \cite{rizzo}.  
Our work is different from these previous works in a few aspects. First of 
all,  we include the possibility that there is a new physics contribution to 
$C_7$ with a new CP violating phase (Im $(\xi_7) \neq 0$).  This necessarily 
calls for studying the direct CP violation in $B\rightarrow X_s \gamma$ as 
advocated by Kagan and Neubert \cite{kn98}, and invalidates the most previous
works on the model-independent determination of $C_{7,9,10}$'s. Secondly we 
include the recent experimental constraint on $R_g$, instead assuming that 
it can be large enough to solve the semileptonic branching ratio problem in 
$B$ decays.  Finally, we assume that the new physics does not introduce any 
new operators with chiralities different from those in the SM, and simply 
modifies the Wilson coefficients of $O_{7,8,9}$. Thus our analysis does not 
consider the left-right symmetric extension of the SM. 

This paper is organized as follows. In Sec.~ II, we give basic formulae
for the relevant physical observables such as $R_{\gamma}$, $R_{ll}$, etc. 
as functions of four real parameters, $| \xi_7 |, {\rm Im}~( \xi_7 ), \chi$ 
and $\chi^{'}$. In Sec. ~III, we present the model-independent numerical 
analysis for both $\chi^{'}=0$ and $\chi^{'} \neq 0$ cases.  
We show the possible ranges of $A_{\rm CP},  
R_{ll},$ etc., when we impose the experimental data on $R_{\gamma}$ 
and $ R_g$. In Sec.~ IV, we discuss explicit SUSY models with gluino-mediated 
FCNC that enjoy  the several assumptions we make in this work. The results of 
this work are summarized in Sec. ~V.  


\section{Relevant Physical Observables}
\subsection{$B\rightarrow X_s \gamma$ and $B\rightarrow X_{sg}$}

In the SM, the branching ratios for $B\rightarrow X_s \gamma$ and 
$B \rightarrow X_{sg}$ are obtained  including the $O(\alpha_s)$ corrections 
and the nonperturbative effects of $b-$quark's Fermi motion inside the $B$ 
meson.  
Relegating the details to the recent works by Kagan and Neubert \cite{kn98}, 
we show the final expressions that will be used in the following : 
\begin{eqnarray}
R_{\gamma} \equiv {B(B\rightarrow X_s \gamma) \over B_{\rm SM} (B\rightarrow 
X_s \gamma)} = 1 + r_1 (\chi) [ {\rm Re} (\xi_7) - 1 ] + r_2 (\chi) [ 
| \xi_7 |^2 - 1 ],
\\
R_{g} \equiv {B(B\rightarrow X_{sg}) \over B_{\rm SM} (B\rightarrow 
X_{sg}) } = 1 + r_3 (\chi) [ {\rm Re} (\xi_7) - 1 ] + r_4 (\chi) [ 
| \xi_7 |^2 - 1 ]. 
\end{eqnarray}
For a real $\chi$ (and $E_{\gamma}^{\rm min}=1.95$ GeV 
for the case of $R_{\gamma}$),
the functions $r_i$'s can be approximated by \cite{kn98}
\begin{eqnarray}
r_1 (\chi) & \approx & 0.46 + 0.020 \chi - 0.0027 \chi^2,
\nonumber   \\
r_2 (\chi) & \approx & 0.11 + 0.025 \chi + 0.0013 \chi^2,
\nonumber   \\
r_3 (\chi) & \approx & 0.43 \chi ( 1 - \chi) + 0.50 \chi,
\nonumber   \\
r_4 (\chi) & \approx & 0.21 \chi^2.   
\end{eqnarray}
The recent CLEO data
\begin{eqnarray}
B(B\rightarrow X_s \gamma) &=& (3.15 \pm 0.35_{\rm stat} \pm 0.32_{\rm syst} 
\pm 0.26_{\rm model}) \times 10^{-4},
~~~\cite{cleo98}
\nonumber     \\
 B(B\rightarrow X_{sg}) & \lesssim & 6.8 \% ~~(90 \% {\rm C.L.}),
~~~\cite{coan}
\end{eqnarray}
and the SM predictions on these decays 
($(3.29 \pm 0.33)\times 10^{-4}$)
imply that 
\begin{eqnarray}
0.77 < R_{\gamma} < 1.15 ~~~(68 \% {\rm C.L.}),
\nonumber  \\
R_{g} \lesssim 6.8/0.2 = 34 ~~~(90 \% {\rm C.L.}).
\end{eqnarray}

\noindent
CP violation in the inclusive $B\rightarrow X_s \gamma$ 
($E_{\gamma}^{\rm min}=1.85$ Gev)
is characterized by CP asymmetry 
\footnote{In this work, the Wilson coefficients without argument
represent those at the scale $\mu=m_b$, whereas those at the $M_W$ scale are
written as $C_i (M_W)$ explicitly.}, 
\begin{eqnarray}
A_{\rm CP}  
&=& {1\over |C_7|^2}~\left\{
1.23 ~{\rm Im} [ C_2 C_7^* ] - 9.52 ~{\rm Im} [ C_8 C_7^* ] 
+ 0.10 ~{\rm Im} [ C_2 C_8^* ] \right\}~(\%)~, \nonumber \\
&=& \frac{A_1 (\chi) ~{\rm Im}(\xi_7)}
  {A_2 (\chi) + A_3 (\chi)~ | \xi_7 |^2 +  A_4 (\chi) ~{\rm Re} (\xi_7)}
~(\%)~,
\end{eqnarray} 
where
\begin{eqnarray}
A_1 (\chi) & \approx & 0.37 - 0.18 \chi,
\nonumber  \\
A_2 (\chi) & \approx & 0.033 - 0.0034 \chi + 0.000085 \chi^2,
\nonumber  \\
A_3 (\chi) & \approx & 0.018 + 0.0025 \chi + 0.000085 \chi^2,
\nonumber  \\
A_4 (\chi) & \approx & 0.049 + 0.00089 \chi - 0.00017 \chi^2.   
\end{eqnarray}

\subsection{$B\rightarrow X_s l^+ l^-$}

Now let us consider the decay $B \rightarrow X_s l^+ l^-$, 
which occurs through the  electroweak penguin 
diagrams and the box diagrams in the SM. 
If there is a new physics beyond the SM, there would be generically dim-6
operators with chiralities different from $O_{9,10}$ shown above
through the  electroweak penguin diagrams and the box diagrams.
Morozumi {\it et al.} considered effects of such new operators 
(10 operators) on the 
branching ratio and the forward-backward asymmetry ($A_{FB}$) in $b 
\rightarrow s l^+ l^-$ \cite{cskim}.  In our opinion, it would be more 
meaningful to consider the effects of modified $C_{7,8}$ 
on the decay $B \rightarrow X_s l^+ l^-$, 
since they are generically given by dim.-5 local operators. 
Especially the effects of $C_7$ is enhanced by $1/s$ factor in the low $s$ 
region (see the third line of  Eq.~(\ref{eq:bsll}) below).  
In any rate, we assume that the new physics does not 
introduce new operators with chiralities different from those in the SM, so 
that we assume that the new physics affects the $b\rightarrow s l^+ l^-$ only 
through modification of the $b\rightarrow s \gamma^*$. Therefore, the Wilson 
coefficients $C_{7,8,9}$ may change (with a new CP-violating phase), and 
$C_{10}$ will not be affected at all in our case. 

The differential branching ratio for $b \rightarrow s l^+ l^- $ is given by
\cite{buras} 
\begin{eqnarray}
{dB (b \rightarrow s l^+ l^-) \over d \hat{s}}
 & = & B( b \rightarrow c e \nu )~ {\alpha^2 \over 4 \pi^2}~ \left| 
{V_{ts}^* V_{tb} \over V_{cb} } \right|^2~{1 \over f_{ph} ( m_c / m_b )
\kappa ( m_c / m_b )} 
~\omega( \hat{s} ) \sqrt{ 1 - {4 m_l^2 \over s} }~
\nonumber  \\
& & \left( | C_9^{\rm eff} |^2 \alpha_1 ( \hat{s},\hat{m_s},\hat{m_l} )
+ | C_{10} |^2 \alpha_2 ( \hat{s},\hat{m_s},\hat{m_l} ) \right. 
\nonumber  \\
& & \left.
+ {4 \over \hat{s}} | C_7 |^2  \alpha_3 ( \hat{s},\hat{m_s},\hat{m_l} )
+ 12 \alpha_4 ( \hat{s},\hat{m_s},\hat{m_l} ) ~{\rm Re}~\{ C_7^* 
C_9^{\rm eff} \} \right),
\label{eq:bsll}
\end{eqnarray}
where all the Wilson coefficients are evaluated at $\mu = m_b$ by the 
renormalization group equations, $\hat{s} = m_{ll}^2 / m_b^2$, the function 
$f_{ph} (x) = 1 - 8 x^2 + 8 x^6 - x^8 - 24 x^4 \ln x$ is the phase space 
factor for the semileptonic $b$ decays, and the function $\kappa(z)$ defined
as 
\begin{equation}
\kappa (z)  = 1 - 2 {\alpha_s 
(m_b) \over 3 \pi}~\left[ ( \pi^2 - {31\over 4} )(1 -z )^2 + {3 \over 2} 
\right]
\end{equation}
is the QCD correction factor thereof. 
The effective Wilson coefficient $C_9^{\rm eff}$ is defined as
\begin{eqnarray}
C_9^{\rm eff} & \equiv & 
C_9 \tilde{\eta} ( \hat{s} ) + Y_{\rm pert} (\hat{s})  
\nonumber  \\
& = & C_9 \tilde{\eta} (\hat{s}) + h(z,\hat{s})
\left( 3 C_1^{(0)} + C_2^{(0)} + 3 C_3^{(0)} + C_4^{(0)} + 3 C_5^{(0)} + 
C_6^{(0)} \right)
\nonumber  \\
& - & {1\over 2}~h(1,\hat{s})~\left( 4 C_3^{(0)} + 4 C_4^{(0)} + 3 C_5^{(0)} 
+ C_6^{(0)} \right)
\nonumber   \\
& - & {1\over 2}~h(0,\hat{s})~\left( C_3^{(0)} + 3 C_4^{(0)} \right) 
+ { 2 \over 9}~\left( 3 C_3^{(0)} + C_4^{(0)} + 3 C_5^{(0)} + C_6^{(0)} 
\right),  
\end{eqnarray}
where $C_i^{(0)}$'s are the Wilson coefficients at $\mu = m_b$ in the leading
logarithmic approximation : 
\begin{equation}
C_j^{(0)} = \sum_{i=1}^8 k_{ji} \eta^{a_i},~~~(j = 1,2,...6)
\end{equation}
with $\eta \equiv \alpha_s (M_W) / \alpha_s (\mu)$, and the numbers $a_i$'s 
and $k_{ji}$'s are given in Table XXVII in Ref.~\cite{rmp}.
The functions $\alpha$'s and $\omega$ are \cite{okada}
\begin{eqnarray}
\alpha_1 (x,y,z) & = & \left( 1 + {2 z \over x} \right) ~\left[ -2 x^2 + 
x ( 1 + y ) + ( 1 - y )^2 \right],
\nonumber  \\
\alpha_2 (x,y,z) & = & \left[ - 2 x^2 + x ( 1+y) + ( 1-y)^2 \right]
+ { 2 z \over x}~\left[ 4 x^2 - 5 ( 1 + y ) x + ( 1 - y )^2 \right]
\nonumber  \\
\alpha_3 (x,y,z) & = & \left( 1 + {2 z \over x} \right)~\left[ - ( 1 + y ) 
x^2 - ( 1 + 14 y + y^2 ) x + 2 ( 1 + y ) ( 1 - y)^2 \right],
\nonumber  \\
\alpha_4 (x,y,z) & = & \left( 1 + {2 z \over x} \right)~\left[ ( 1 - y )^2 
- ( 1 + y ) x \right],
\nonumber  \\
\omega (\hat{s}) & = & \sqrt{ \left[ \hat{s} - ( 1 + \hat{m_s} )^2 \right]~
\left[ \hat{s} - ( 1 - \hat{m_s} )^2 \right] },
\end{eqnarray} 
with $\hat{m_s} \equiv m_s^2 / m_b^2 \equiv 0$. 
And ${\tilde\eta}({\hat s})$ is given by \cite{buras}
\begin{eqnarray}
{\tilde\eta}({\hat s})&=&1+\frac{\alpha_s(\mu)}{\pi}{\tilde\omega}({\hat s}),
\nonumber \\
{\tilde\omega}({\hat s})&=&
-\frac{2}{9}\pi^2-\frac{4}{3}{\rm Li}_2({\hat s})-\frac{2}{3}\ln{\hat
s}\ln(1-{\hat s})-\frac{5+4{\hat s}}{3(1+2{\hat s})}\ln(1-{\hat s})
\nonumber \\
&&-\frac{2{\hat s}(1+{\hat s})(1-2{\hat s})}{3(1-{\hat s})^2(1+2{\hat
s})}\ln{\hat s}+\frac{5+9{\hat s}-6{\hat s}^2}{6(1-{\hat s})(1+2{\hat s})}.
\end{eqnarray}
The function $Y_{\rm pert}(\hat{s})$ represents the $O(\alpha_s)$ 
corrections of the matrix elements, whose explicit form can be found at 
Ref.~\cite{buras}. 
The new physics contributions can induce $b\rightarrow s q \bar{q}$
through $b\rightarrow s g^* \rightarrow s q \bar{q}$. This will modifies the 
Wilson coefficients $C_{i=1-6}$'s, 
whose effects can be seen in 
the direct CP violation in the $B$ decay amplitude. However these will not 
affect $b\rightarrow s \gamma$ and $b\rightarrow s l^+ l^-$ at the order we are
working on. For the realistic prediction, one also has to include the 
long distance contribution 
through $b \rightarrow (J/\psi,\psi^{'}) + s$ followed by
($J/\psi,\psi^{'}) \rightarrow l^+ l^-$. 
This can be taken into account by adding to
the perturbative $Y_{\rm pert}(\hat{s})$ 
the resonance contributions \cite{morozumi} :
\begin{equation}
Y_{\rm res} ( \hat{s} ) = \tilde{\kappa}~{3 \pi \over \alpha^2}~
\sum_{i=J/\psi, \psi^{'}}~{M_i \Gamma ( i \rightarrow l^+ l^-) / m_b^2 
\over \hat{s} - M_i^2 / m_b^2 + i M_i \Gamma_i / m_b^2 } \quad , 
\end{equation}
with $\tilde{\kappa} = -1$.
To avoid the large contributions from the $J/\psi$ and $\psi^{'}$ resonances,
we consider the following two regions:
the low $\hat{s}$ region, $1$ GeV$^2 < \hat{s}~ m_b^2 < 8~{\rm GeV}^2$ for 
$b \rightarrow s e^+ e^-$ case,
and the high $\hat{s}$ region, 
$0.6<\hat{s}<1$ for $b \rightarrow s \tau^+ \tau^-$.

Using these informations, it is straightforward to evaluate $R_{ll}$ :
\begin{eqnarray}
R_{ll} \equiv {B(B\rightarrow X_s l^+ l^-) \over B_{\rm SM} (B\rightarrow
X_s l^+ l^-)} \quad .
\end{eqnarray}
For the decay $B\rightarrow X_s e^+ e^-$,
\begin{eqnarray}
B(B\rightarrow X_s e^+ e^-) &=& B_1 + B_2 |C_9|^2 + B_3 {\rm Re} (C_9)
+ B_4 {\rm Im} (C_9) \nonumber \\
&+& B_5 |C_7|^2 + B_6 {\rm Re} (C_7^* C_9)
+ B_7 {\rm Re} (C_7) + B_8 {\rm Im} (C_7) \quad ,
\end{eqnarray}
with
\begin{eqnarray}
B_1 & \approx & 1.89,~
B_2  \approx  0.07,~
B_3  \approx  0.19,~
B_4  \approx  0.007,~
\nonumber \\
B_5 & \approx & 4.07,~
B_6  \approx  0.68,~
B_7  \approx  0.87,~
B_8  \approx  0.034~~~~~  (\times 10^{-6}).
\end{eqnarray}
For the decay $B\rightarrow X_s \tau^+ \tau^-$,
\begin{eqnarray}
B(B\rightarrow X_s \tau^+ \tau^-) &=& D_1 + D_2 |C_9|^2 + D_3 {\rm Re} (C_9)
+ D_4 {\rm Im} (C_9) \nonumber \\
&+& D_5 |C_7|^2 + D_6 {\rm Re} (C_7^* C_9)
+ D_7 {\rm Re} (C_7) + D_8 {\rm Im} (C_7),
\end{eqnarray}
with
\begin{eqnarray}
D_1 & \approx & 12.6,
D_2  \approx  0.87,
D_3  \approx  -1.51,
D_4  \approx  1.66,
\nonumber \\
D_5 & \approx & 6.24,
D_6  \approx  4.58,
D_7  \approx  -4.21,
D_8  \approx  4.32~~~~~  (\times 10^{-8}).
\end{eqnarray}

Another interesting observable at B factories is the forward-backward
asymmetry of the lepton in the center of mass frame of the lepton pair :
\begin{eqnarray}
A_{FB} & \equiv & 
{{ [ \int_{0}^1 d(\cos\theta) - \int_{-1}^0 d(\cos\theta) ] 
~d^2B/d\hat{s} d\cos\theta } \over 
{[ \int_{0}^1 d(\cos\theta) + \int_{-1}^0 d(\cos\theta) ] 
~d^2B/d\hat{s} d\cos\theta } }
\nonumber  \\
& = & - {3 \omega(\hat{s}) \sqrt{1 - 4 \hat{m_l}^2 / \hat{s} } ~C_{10}
{\rm Re}~\left\{ \hat{s} [ C_9 + Y ( \hat{s} ) ] + 2 C_7 \right\}
\over \left\{ | C_9 + Y(\hat{s}) |^2 \alpha_1 + C_{10}^2 \alpha_2
+ (4/\hat{s}) C_7^2 \alpha_3 + 12 \alpha_4 {\rm Re} C_7 [ C_9 + Y(\hat{s})
] \right\}  },
\end{eqnarray}
where $\theta$ is the angle between the positively charged lepton and the 
$B$ flight direction in the rest frame of the dilepton system.  
For the decay $B\rightarrow X_s e^+ e^-$, the integrated forward-backward 
asymmetry is given by 
\begin{equation}
A_{FB} (ee) =  { E_1 + E_2 {\rm Re} (C_9) + E_3 {\rm Re} (C_7) \over
B(B\rightarrow X_s e^+ e^-) } \quad ,
\end{equation}
where
\begin{eqnarray}
E_1  \approx  0.21,~
E_2  \approx  0.14,~
E_3  \approx  1.69~~~~~~(\times 10^{-6}) \quad .
\end{eqnarray}
For the decay $B\rightarrow X_s \tau^+ \tau^-$,
\begin{equation}
A_{FB} (\tau \tau) =  { F_1 + F_2 {\rm Re} (C_9) + F_3 {\rm Re} (C_7) \over
B(B\rightarrow X_s \tau^+ \tau^-) },
\end{equation}
where
\begin{eqnarray}
F_1  \approx  -0.86,
F_2  \approx  1.26,
F_3  \approx  3.66~~~~~~(\times 10^{-8}).
\end{eqnarray}
The last observable we discuss is the tau polarization asymmetry
$P_{\tau} (\hat{s}) $ in $B\rightarrow X_s \tau^+ \tau^-$ defined as
\cite{hewett}
\begin{eqnarray}
P_{\tau} ( \hat{s} ) & \equiv &
{ {dB\over d\hat{s}}_{\lambda=-1} - {dB\over d\hat{s}}_{\lambda=+1}
\over  {dB\over d\hat{s}}_{\lambda=-1} + {dB\over d\hat{s}}_{\lambda=+1} }
\nonumber        \\
& = & { -2 \omega(\hat{s}) \sqrt{1 - 4 \hat{m_l}^2/\hat{s}}  C_{10}~
{\rm Re} [ ( 1 + 2 \hat{s}) \{ C_{9} + Y(\hat{s}) \} + 6 C_7 ] 
\over \left\{ | C_9 + Y(\hat{s}) |^2 \alpha_1 + C_{10}^2 \alpha_2
+ (4/\hat{s}) C_7^2 \alpha_3 + 12 \alpha_4 {\rm Re} C_7 [ C_9 + Y(\hat{s})
] \right\}  } \quad .
\end{eqnarray}
The integrated tau polarization asymmetry $P_{\tau} $ can be expressed as
\begin{equation}
P_{\tau} = {T_1 + T_2 {\rm Re} (C_9) + T_3 {\rm Re} (C_7) \over
B(B \rightarrow X_s \tau^+ \tau^-)} \quad ,
\end{equation}
where
\begin{eqnarray}
T_1 \approx -1.99,~
T_2 \approx  2.83,~
T_3 \approx  7.31 ~~~~(\times 10^{-8}) \quad .
\end{eqnarray}
Since $B$ decays into the tau pair probes high $m_{\tau\tau} ( > 3.554$ GeV) 
region, the observable $P_{\tau}$ is sensitive to the deviation of 
$C_{9}$ from their SM values which dominates the $B\rightarrow X_s \tau^+
\tau^-$ at high $s$ region. 

\section{Model-independent analysis}

Now we are ready to do a model-independent analysis using the formulae 
obtained in the previous section.  There are two different cases depending 
on $\chi^{'} = 0$ or not.  In principle, any new physics contributing to
magnetic form factor in $b\rightarrow s\gamma$ may affect the electric form 
factor as well. Therefore one would expect generically $\chi^{'} \neq 0$.
However this needs not be necessarily true as discussed in the next section 
(the case (i)). So we discuss $\chi^{'} = 0$ and $\chi^{'} \neq 0$ separately
in this section.

Our strategy is the following : impose the experimental data on $R_{\gamma}$ 
and $R_g$ :
\begin{itemize}
\item {\bf E1 } : 
$0.77 < R_{\gamma} < 1.15$ as in Ref.~\cite{cleo98} 
\item {\bf E2} :
$R_g < ( 6.8 \% / 0.2\% ) = 34$ \cite{coan}
\end{itemize}
For given $\chi$ and $\chi^{'}$, these constraints (E1) and (E2) determine 
the allowed region in the complex $\xi_7$ plane. Then, in the allowed $\xi_7$
plane, one can calculate other physical observables, $A_{CP}, R_{ll}, 
A_{FB} (b\rightarrow sll)$ and $P_{\tau}$.  
Because the number of observables are greater than the number of unknown
parameters (one complex number $\xi_7$ and two real numbers $\chi$ and 
$\chi^{'}$), one can overconstrain these 4 real parameters. If there is no
consistent solution, there would be a few possibilities : $\chi$ and/or 
$\chi^{'}$ may be complex, $C_{10}$ is modified by new physics effects,
or one has to enlarge the operator basis by including operators with 
different chiralities from those in the SM, as in Ref.~\cite{cskim}. 

\begin{figure}[ht]
\hspace*{-1.0 truein}
\psfig{figure=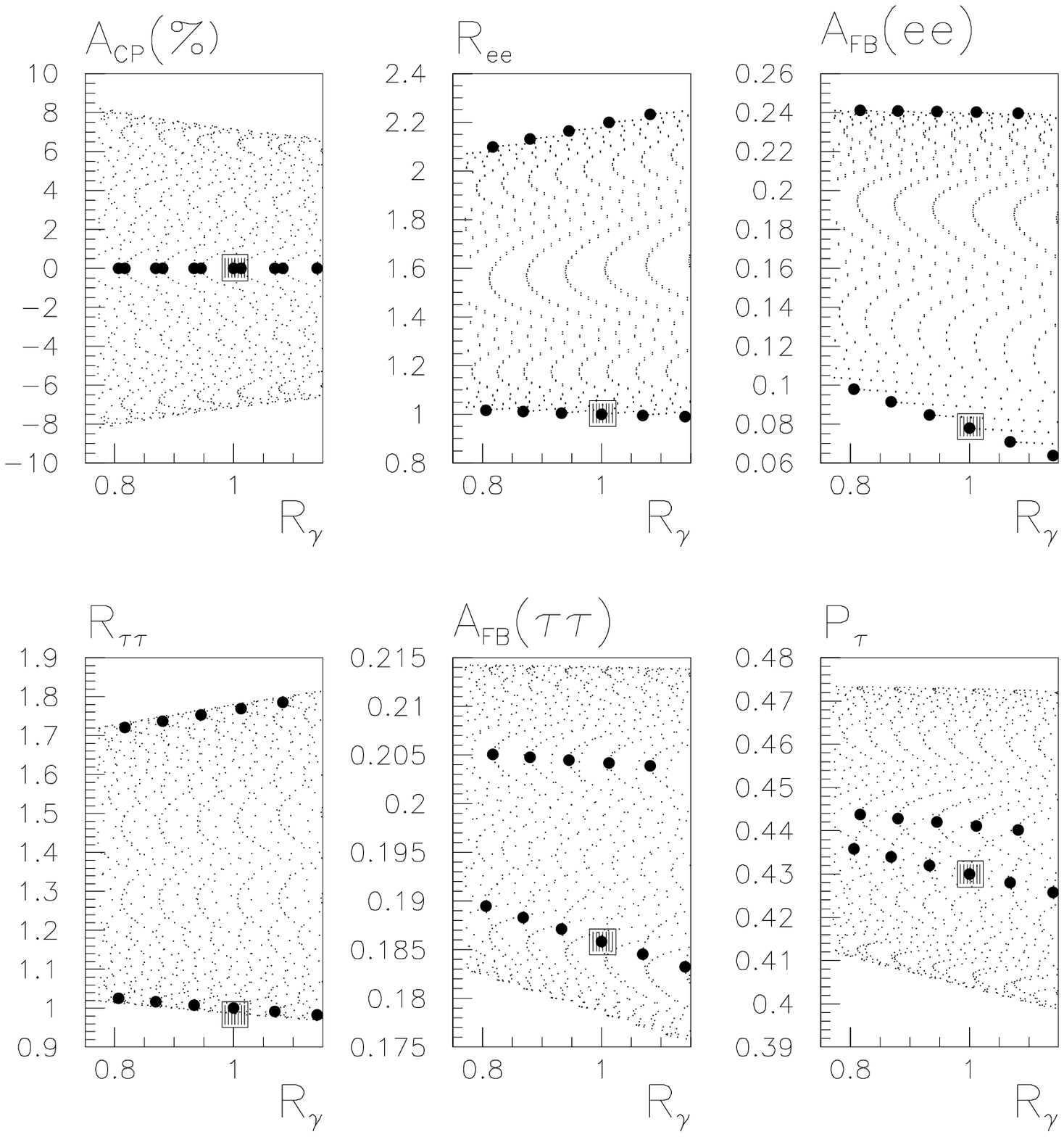}
\caption{Scattered plots for $A_{CP}, R_{ee}, 
A_{FB}(ee), R_{\tau\tau}, A_{FB}(\tau\tau)$ and $P_{\tau}$ as functions of 
$R_{\gamma}$ for $\chi = \chi^{'} = 0$. 
The SM values are marked as a filled square.
}
\label{fig_ch0}
\end{figure}

Let us first consider the case with $\chi^{'} = 0$. 
In Fig.~\ref{fig_ch0}, we show the scattered plots of various observables 
as functions of $R_{\gamma}$ for $\chi=0$. The SM case is denoted by a 
square, possible values in our model are represented by dots,  
whereas the filled circles represent the case where there is no  new CP 
violating phase, namely ${\rm Im} \xi_7=0$, but ${\rm Re} \xi_7 \neq 0$.  
Implications of these figures are clear. For example, the CP asymmetry in 
$b\rightarrow s\gamma$ cannot be larger than $\sim \pm 8 \%$ if $\chi = 0$, 
and $R_{ee}$ can be anywhere between 0.98 to 2.2.  For comparison, let us
discuss the minimal SUGRA model with universal soft mass terms at GUT scale, 
in which typical values of $\chi$ and $\chi^{'}$ are $\chi \sim 1$ and 
$\chi^{'} \lesssim  0.05, \chi \approx 0.05$ respectively \cite{okada}. 
Therefore, the predictions in the minimal SUGRA model are very close to 
the dots in Fig.~\ref{fig_ch0}.
Namely, in the SUGRA case, there are two bands for the possible 
$R_{ee}$ for a given $R_{\gamma}$, whereas in our case, $R_{ee}$ can be 
anywhere in between becasue of the presence of a new CP-violating phase 
given by ${\rm Arctan} ( {\rm Im} \xi_7 / {\rm Re} \xi_7 )$.

\begin{figure}[ht]
\hspace*{-1.0 truein}
\psfig{figure=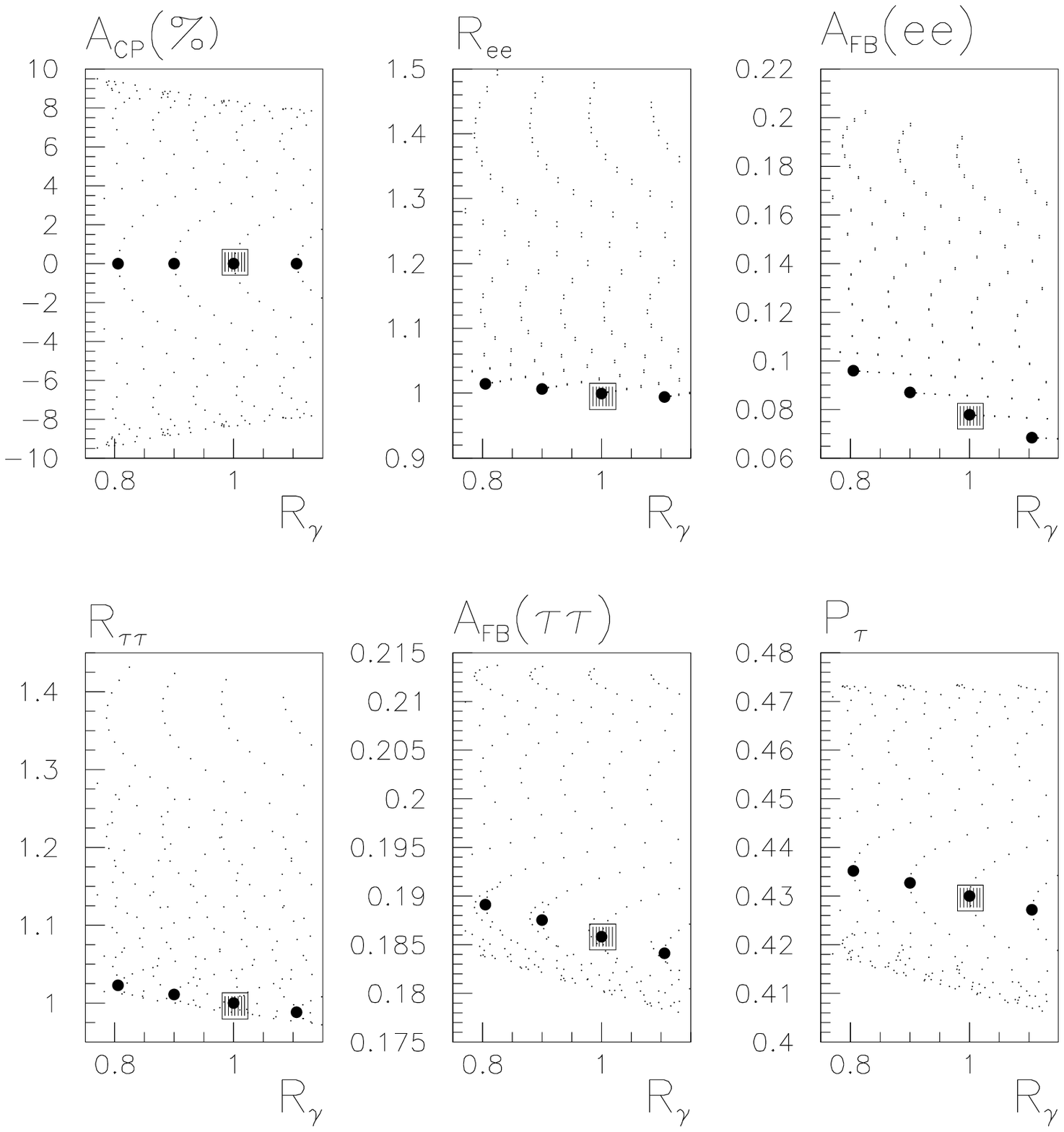}
\caption{The same as Fig.~1 for $\chi = 5$ and $\chi^{'} = 0$. 
The SM values are marked as a filled square.
}
\label{fig_ch5}
\end{figure}

\begin{figure}[ht]
\hspace*{-1.0 truein}
\psfig{figure=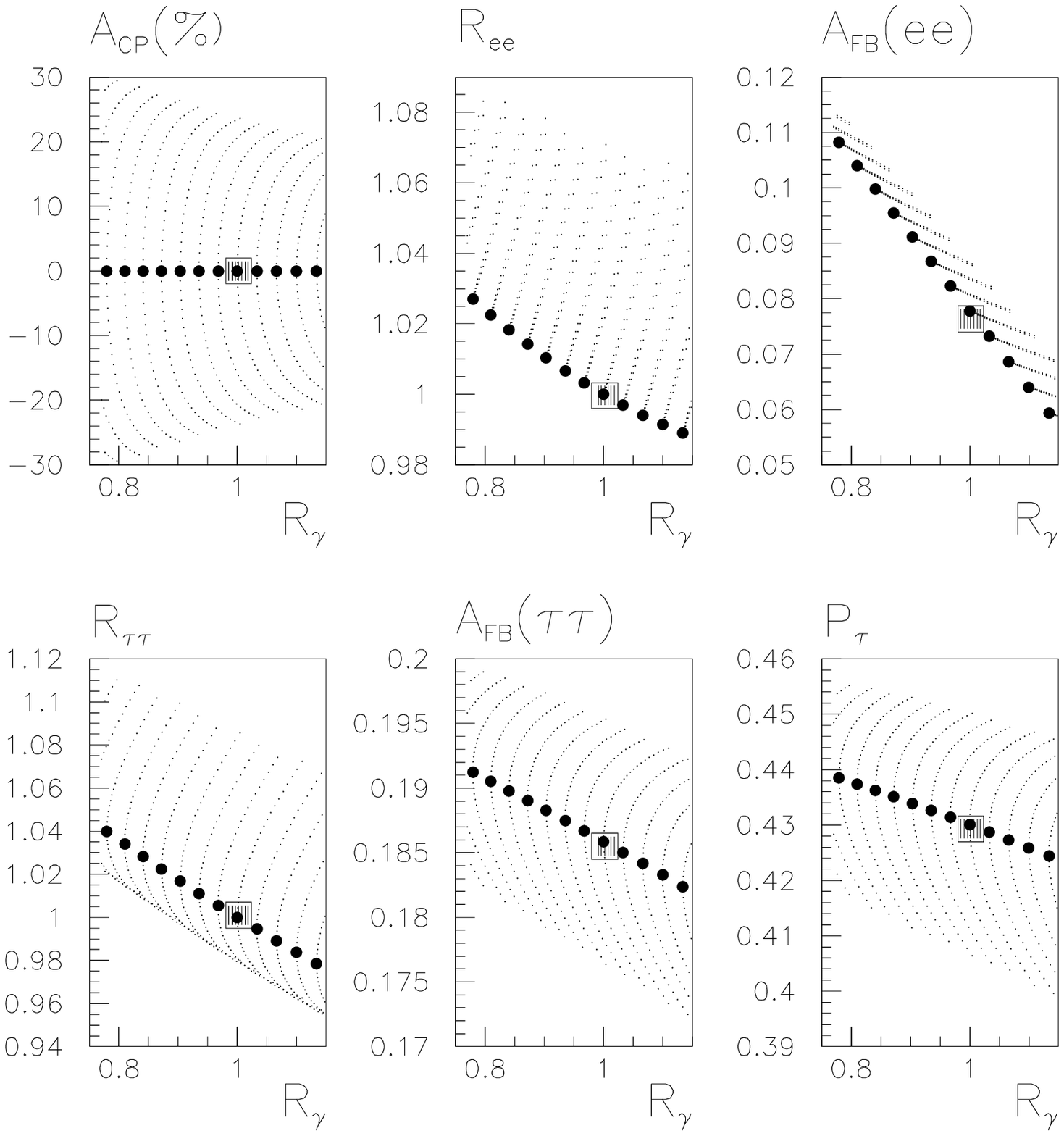}
\caption{The same as Fig.~1 for $\chi = - 5$ and $\chi^{'} = 0$. 
The SM values are marked as a filled square.}
\label{fig_ch-5}
\end{figure}

In Figs.~\ref{fig_ch5} and \ref{fig_ch-5}, we show similar plots for 
$\chi = 5$ and $\chi = -5$, respectively. This choice of $\chi$ covers a 
large class of new physics as discussed in Ref.~\cite{kn98}. 
Implications of these figures are almost the same as Fig.~1, except that 
there is now rather strong constraint from 
$B \rightarrow X_{s g}$ (E2).  In this case we can have larger direct CP
violation in $B\rightarrow X_s \gamma$ upto $10-30 \%$.  Also the (E2) 
constraint removes substantial  parts of available $R_{ee}, A_{FB}(ee)$ 
and $R_{\tau\tau}$ as shown in Fig.~\ref{fig_ch5} ($\chi = 5$), 
compared to Fig.~\ref{fig_ch0} where the constraint (E2) was not imposed. 
This effect is much more prominent for
negative $\chi$ as shown in Fig.~\ref{fig_ch-5} ($\chi=-5$).  For example, 
the $A_{FB}(ee) - R_{\gamma}$ correlation is almost identical to the case
with vanishing new phase ${\rm Im} \xi_7 = 0$.  From Figs.
\ref{fig_ch0}--\ref{fig_ch-5}, it is clear that the existence of a new CP
violating phase not only can generate a large CP asymmetry  in $b\rightarrow 
s \gamma$, but can it also induce quite a lot deviations of various 
observables in $b\rightarrow s l^+ l^-$ for $l=e,\mu$ and $\tau$.  For 
$\chi=0$ and $\chi = 5$, deviations of the observables  $A_{\rm CP}, R_{ee}, 
A_{FB}(ee), R_{\tau \tau}$ from their SM values can be large enough that 
they can be clearly observed at future B factories, whereas deviations of 
other observables $A_{FB} (\tau\tau)$ and $P_{\tau}$ from their SM values 
are rather small that it would be very difficult to measure them.  For 
$\chi=-5$, only $A_{\rm CP}$ and $A_{\rm FB} (ee)$ shows substantial 
deviations from the SM values because of the (E2) constraint again. 
If the experimental data on $A_{FB}(\tau\tau)$ and $P_{\tau} $ show large 
deviations from their SM values, it would indicate that $\chi$ and/or 
$\chi^{'}$ are complex, or some new physics contributes to $C_{10}$ (with a 
possibly new CP violating phase), and/or even generates $O_{7(8),R}$ and 
possibly other dimension-6 $bsll$ operators with different chiralities from 
$O_{9,10}$ in the SM.

\begin{figure}[ht]
\hspace*{-1.0 truein}
\psfig{figure=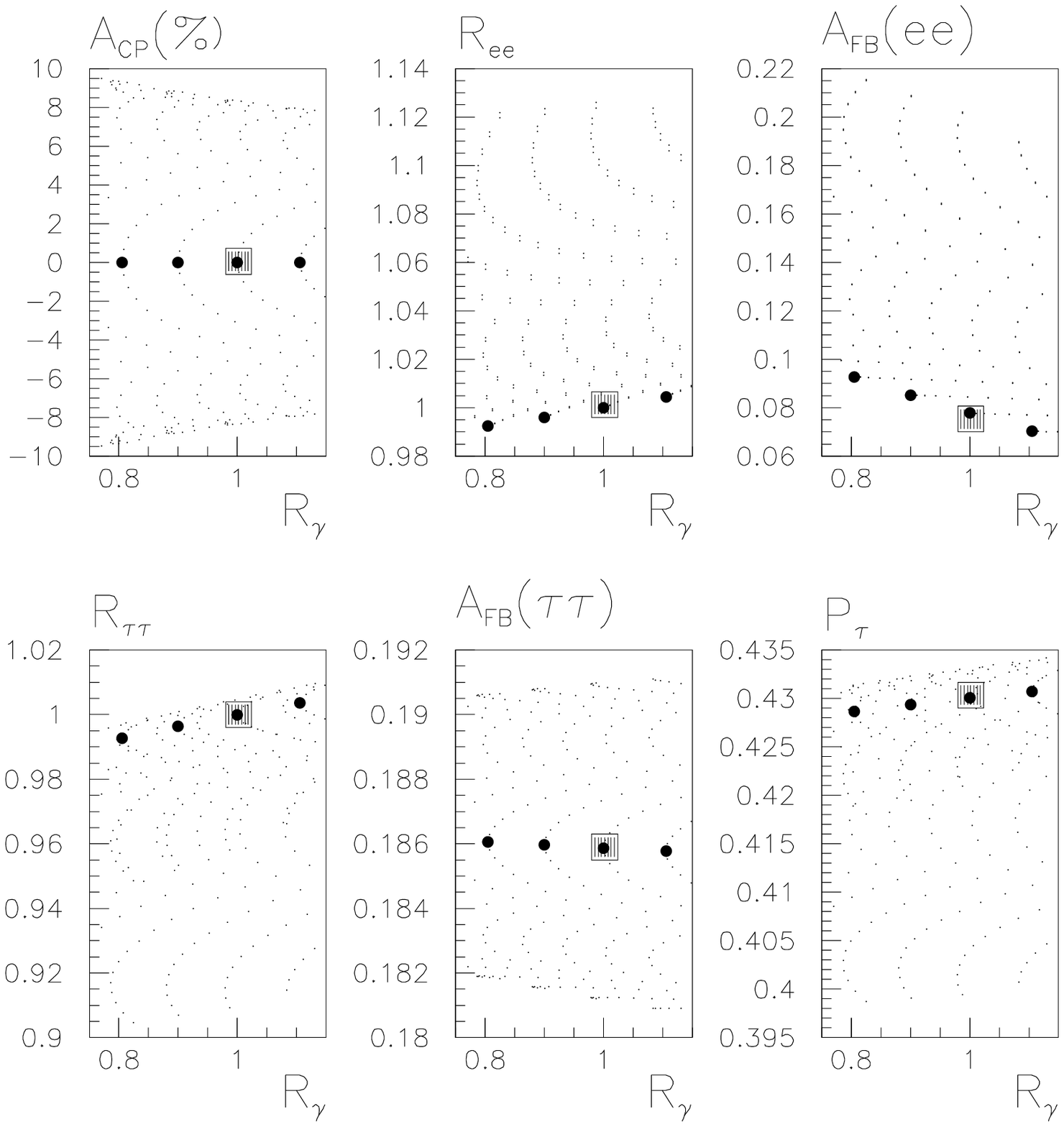}
\caption{The same as Fig.~1 for $\chi = 5$ and $\chi^{'} = 0.3$. 
The SM values are marked as a filled square.}
\label{fig_ch5pr03}
\end{figure}

The nonvanishing $\chi^{'}$ does not affect $b\rightarrow s \gamma$ and 
$b\rightarrow s g$ so that the allowed region in the complex $\xi_7$ plane 
remains the same as before, for a given $\chi$. However, it does change 
the observables related with $B \rightarrow X_s l^+ l^-$, and  we show them
in Fig.~\ref{fig_ch5pr03} for $(\chi, \chi^{'}) = (5,0.3)$, where we chose 
$\chi^{'}=0.3$ that is typical in the gluino-mediated SUSY models considered
in the next section. The $R_{ee}, R_{\tau\tau}, P_{\tau}$  dependence on 
$R_{\gamma}$ differ from those in Fig.~\ref{fig_ch5pr03}, and the possible 
deviations of these observables from their SM values are smaller if $\chi^{'}
= 0.3$. If there is no new CP violating phase, the differences are so tiny 
that one may not be able to distinguish two cases in practice.

The message of this model-independent study is that the previous methods
\cite{cskim}-\cite{rizzo} has to be enlarged to include a new observable 
$A_{\rm CP}$ that could be sensitive to a new CP violating phase.  In the 
presence of such a new phase, simple correlations among various observables 
in $B \rightarrow X_s \gamma$ and $B\rightarrow X_s l^+ l^-$ (namely, 
correlations among $R_{\gamma}, R_{ll}, A_{FB}(ll)$  with ($l = e, \mu, 
\tau$) and $P_{\tau}$ that are represented as thick  dots in Figs.~1-4 
simply disappear,
and there is no more apparent correlations among these observables. Still
one can perform a global analysis as before using the formulae given in the
previous sections, including the observable  $A_{\rm CP}$. 
This will provide additional information and one can overconstrain
four real parameters, ${\rm Re} \xi_7,{\rm Im} \xi_7, \chi$ and $\chi^{'}$.  
If there is no consistent solution for these four real parameters, one has to 
consider the possibility that $\chi$ and/or $\chi^{'}$ are also complex. 
In this case one may be able to determine the Wilson coefficients, 
if one can measure all the observables related with $B\rightarrow X_s l^+ 
l^-$. This task will be possible, only after B factories accumulate the 
data for the first several years. Or one might have to consider the modified 
$C_{10}$ and new operators that are not possible in the SM.   

\section{Specific SUSY models with gluino-mediated FCNC }

\subsection{Models}

In the previous section, we presented model-independent analysis of physics
related  with $C_{7,8,9}$ assuming there is a new CP-violating phase and 
both $\chi$ and $\chi^{'}$ are real.  In this section, we wish to present  
specific models that satisfy  such assumptions. 
Let us consider the FCNC in generalized SUSY models, in which squark mass 
matrices are nondiagonal in the basis where fermion mass matrices are diagonal. 
In this case, there can be a potentially  important contributions to the 
FCNC processes and CP violation that arise from  flavor changing 
$f-\tilde{f}-\tilde{g}$ vertices \cite{masiero}. The sources of SUSY FCNC 
are the nondiagonal $(M_{LL}^{\tilde{q}})^2, (M_{LR}^{\tilde{q}})^2$ and 
$(M_{RR}^{\tilde{q}})^2$.  Different SUSY breaking models have different 
patterns/hierarchies for the flavor mixings in the squark mass matrices. 
Since we study the new physics contributions to $C_{7,8,9}$, the Wilson 
coefficients of the operators already present in the SM in this work, 
we will consider only two cases : 
(i) the $(LR)$ mixing dominating and (ii) the $(LL)$ mixing 
dominating cases.  There are some models in the literature which fall into 
these  two categories.  As discussed below, the case (i) does not contribute 
to $C_9$ so that $\xi_9 = 1$ (or,  $\chi^{'}=0$). On the other hand, the case 
(ii) contributes to $C_9$ as well as to $\xi_7$ and $\xi_8$. 
Also, there would be generically other contributions from 
$H^- - t, \chi^{-} - \tilde{t}$ and $\chi^0 - \tilde{d}_k$ loops
\footnote{After we submitted this paper, there appeared works which considered
these effects in the most general MSSM \cite{baek}, in the minimal 
supergravity scenario \cite{keum} and its modified versions \cite{aoki}.}.
If these loop effects are competent with the gluino-mediated loop effects
we consider in the following, then our assumption that both $\chi$ and 
$\chi^{'}$ are real would not be true any longer.   
In the following, we assume that these (SUSY) electroweak loops  are indeed 
negligible compared to the gluino mediated FCNC loop amplitudes. 
The latter is enhanced by $\alpha_s / ( G_F m_W^2 )$, as usually assumed. 
However there is a suppression factor in the latter case, the mixing angle in 
the squark sector given by $\Gamma_{GL}^d$ (or, $( \delta_{23}^d )_{LL}$ 
in the mass insertion approximation). 
Also the heavy squark-gluino loops will be suppressed compared 
to the charged Higgs - top, chargino - stop and neutralino - down squarks, 
unless all the SUSY particles have similar masses so that squark and gluinos 
are not too heavy. 
So one has to keep in mind that our assumption may break down 
for too small mixing angle in the squark sector or too heavy squark/gluino. 
With this caveat in mind, new physics contributions considered here  
depend on only one new phase so that  $\chi$ and $\chi^{'}$ are real, 
as assumed in the previous section.

In order to estimate the $\xi_{7,8,9}$ in the generalized SUSY models with
gluino-mediated FCNC, we consider both the vertex mixing 
(VM) method and the mass insertion approximation (MIA). 
The latter approximation is good,  when squarks are almost degenarate. 
The corresponding expressions can be obtained from the former expressions 
by taking a suitable expansion in 
$\Delta \tilde{m}^2 \equiv ( ( M^{\tilde{d}} )^2 - \tilde{m}^2)$, 
where $\tilde{m}$ is a suitable  average mass of almost degenerate squarks. 
On the other hand, in the scenario in which the SUSY FCNC and SUSY CP problem
are solved by decoupling of the (nearly degenerate) first two generation 
squarks such as in the effective SUSY models, there is a large hierarchy 
between  the first two and the third squarks so that the MIA is no longer a 
good approximation. In such case, we have to resort to the VM method.

The full expressions for the Wilson coefficients $C_{7,8,9}$ due to the FCNC 
gluino exchange diagrams are 
\cite{okada} 
\begin{eqnarray}
C_{7}^{SUSY} (M_W) 
& = & -{8 \pi \alpha_s  \over 9 \sqrt{2} G_F  m_{\tilde{g}}^2 
\lambda_t} ~\sum_{I=1}^6 ~x_I ( \Gamma_{GL}^{d \dagger} )_{2I}
\nonumber  \\
& \times & \left[  ( \Gamma_{GL}^d )_{I3} f_2 (x_I) + 
( \Gamma_{GR}^d )_{I3} { m_{\tilde{g}} \over m_b} f_4 (x_I) 
\right] ,
\nonumber 
\\
C_{8}^{SUSY} (M_W) 
& = & -{\pi \alpha_s  \over  \sqrt{2} G_F  m_{\tilde{g}}^2 
\lambda_t}~\sum_{I=1}^6 ~x_I ( \Gamma_{GL}^{d \dagger} )_{2I} 
\nonumber  
\\
& \times & 
\left[  ( \Gamma_{GL}^d )_{I3} \left\{  3 f_1 (x_I) + {1\over 3} f_2 (x_I) 
\right\} + ( \Gamma_{GR}^d )_{I3} {m_{\tilde{g}} \over m_b} \left\{ 
3 f_3 (x_I) + {1 \over 3} f_4 (x_I) \right\} \right] ,
\nonumber
\\
C_{9}^{SUSY} (M_W) 
& = & {16 \pi \alpha_s  \over 9 \sqrt{2} G_F  m_{\tilde{g}}^2 
\lambda_t} ~\sum_{I=1}^6 ~x_I ( \Gamma_{GL}^{d \dagger} )_{2I} 
( \Gamma_{GL}^d )_{I3} f_6 (x_I) ,
\end{eqnarray}
where $x_I \equiv m_{\tilde{g}}^2 / m_{\tilde{d_I}}^2$. $\Gamma_{GL}^d$ 
and $\Gamma_{GR}^d$ determine the $\tilde{g}-\tilde{d}_i-d_j$ vertices 
as follows :
\begin{equation}
{\cal L} = -g_s \sqrt{2} (T^a)_{\alpha \beta} \overline{\tilde{G}^a} 
~\left[ ( \Gamma_{GL}^d )_{Ij} P_L + ( \Gamma_{GR}^d )_{Ij} P_R \right]
d_{j\alpha} \tilde{d}_{I\beta}^{*} ,  
\end{equation}
with $I,J=1,2,...6$ and $i,j = 1,2,3$.
They are related with the mixing matrix elements diagonalizing the 
down-squark mass matrix via $\tilde{d}_I = (\tilde{U}_D)_{IJ} (\tilde{d}_L,
\tilde{d}_R )_J^{T}$, $\tilde{U}_D M^2_{\tilde{d}} \tilde{U}_D^{\dagger} =
{\rm diagonal}$, with the following identification :
\begin{equation}
( \Gamma_{GL}^d )_{Ij} = ( \tilde{U}_D )_{Ij}, ~~~~~~~~
( \Gamma_{GR}^d )_{Ij} = -( \tilde{U}_D )_{I,j+3}.
\end{equation} 
The functions $f_i$'s are given by \cite{okada}
\begin{eqnarray}
f_1 (x) & = & {x^3 - 6 x^2 + 3 x + 2 + 6 x \ln x \over 12 (x-1)^4} ,
\nonumber   
\\
f_2 (x) & = & {2 x^3 + 3 x^2 - 6 x + 1 - 6 x^2 \ln x \over 12 (x-1)^4} ,
\nonumber 
\\
f_3 (x) & = & {x^2 - 4 x + 3 + 2 \ln x \over 2 (x-1)^3} ,
\nonumber  
\\
f_4 (x) & = & {x^2 - 1 - 2 x \ln x \over 2 (x-1)^3} ,
\nonumber 
\\
f_6 (x) & = & {-11 x^3 + 18 x^2 - 9 x + 2 + 6 x^3 \ln x \over 36 (x-1)^4} .
\end{eqnarray}
The corresponding expressions in the MIA is obtained from the above 
expressions by making a Taylor expansion around $ \tilde{m}$ as follows :
$x_i = ( \tilde{m}^2 + \Delta m_{\tilde{d_i}}^2 ) / \tilde{m}^2$ and using 
the unitarity condition for $\tilde{U}_D$. 
This way one can recover the results in Ref.~\cite{masiero}.  For 
completeness, we present the resulting expressions below :
\begin{eqnarray}
C_{7}^{SUSY} (M_W) 
& = & {8 \pi \alpha_s  \over 9 \sqrt{2} G_F  \tilde{m}^2 
\lambda_t} ~\left[ \left( \delta_{23}^d \right)_{LL} M_3 (x) +  
\left( \delta_{23}^d \right)_{LR} {m_{\tilde{g}} \over m_b} M_1 (x) \right] ,
\nonumber 
\\
C_{8}^{SUSY} ( M_W) 
& = & { \pi \alpha_s  \over  \sqrt{2} G_F  \tilde{m}^2 
\lambda_t}~\left[ \left( \delta_{23}^d \right)_{LL} \left( {1\over 3} M_3 (x) 
+ 3 M_4 (x) \right) +  \left( \delta_{23}^d \right)_{LR} {m_{\tilde{g}} \over 
m_b} \left( {1\over 3} M_1 (x) + 3 M_2 (x) \right) \right],
\nonumber
\\
C_{9}^{SUSY} ( M_W ) & = & 
{16 \pi \alpha_s  \over 9 \sqrt{2}  G_F  \tilde{m}^2 
\lambda_t} ~\left( \delta_{23}^d \right)_{LL} P_1 (x) .
\end{eqnarray}
The functions $M_{1,3}(x)$ and $P_1 (x)$ are defined as 
\begin{eqnarray}
M_1 (x) & = & {1 + 4 x - 5x^2 + ( 4 x + 2 x^2 ) \ln x \over 2 (1-x)^4}, 
\nonumber   \\
M_2 (x) & = & -x^2 { 5 - 4 x - x^2 + ( 2 + 4 x) \ln x \over
2 ( 1 - x)^4 },
\nonumber  \\
M_3 (x) & = & {-1 + 9 x + 9 x^2 - 17 x^3 + ( 18 x^2 + 6 x^3 ) \ln x 
\over 12 (x-1)^5},
\nonumber   \\
M_4 (x) & = & {-1 - 9 x + 9 x^2 + x^3 - 6 (x + x^2 ) \ln x \over 
6 ( x - 1)^5},
\nonumber   \\
P_1 (x) & = & { 1 - 6 x + 18 x^2 - 10 x^3 - 3 x^4 + 12 x^3 \ln x 
\over 18 (x-1)^5 }.
\end{eqnarray}

In order to estimate the $\xi_7, \chi$ and $\chi^{'}$, we assume that the
$(23)$ mixing is the same order of the corresponding CKM matrix element 
with an unknown new phase $\phi \sim O(1)$.  For example, $\delta_{23} \sim 
|\lambda_t| \times e^{i \phi}=A\lambda^2e^{i\phi}$ 
with $\lambda = \sin \theta_c = 0.22$ for 
both cases (i) and (ii), and similarly for $\Gamma_{GL,GR}^d$.
Then it is clear that $\chi$ and $\chi^{'}$ are real in the MIA both in the 
cases (i) and (ii).  In case of the VM approximation, the relevant model is 
the effective SUSY model where only the third family squark can be lighter 
than $\sim 1 $ TeV so that $x_1, x_2  \approx 0$ and we may keep only terms 
proportional to $x_3$ in the summation over $I=1-6$ in Eqs.~(40). 
Then, the $\chi$ and $\chi^{'}$ are real again, as assumed in the previous 
section.  Finally, in the following subsection, we will consider only two 
observables $A_{\rm CP}$ and $R_{ee}$ for simplicity  among several 
observables considered in the previous section. These two observables will
be sufficient for us to find out the generic features considered in the 
previous  section in the specific SUSY models with gluino mediated FCNC.


    
\subsection{Case (i) : $(LR)$ insertion}

Let us first discuss the case (i) : $(LR)$ insertion. Since the flavor 
changing $(LR)$ mixing terms are not generated by SUSY breaking in the limit 
of vanishing Yukawa couplings, they are proportional to the corresponding 
Yukawa couplings. Therefore, the mass insertion approximation is always
appropriate, and we consider the $(LR)$ insertion only in the MIA. From 
Eqs.~(44), one  gets 
\begin{eqnarray}
\xi_7 & = & 1 + {1\over C_{7L}^{\rm SM} (m_W) }~
{8 \pi \alpha_s \over 9 \sqrt{2} G_F \tilde{m}^2 \lambda_t} 
( \delta_{23} )_{LR} {m_{\tilde{g}} \over m_b} M_1 (x) ,
\nonumber   \\
\chi & = & {9~C_{7L}^{\rm SM}(M_W)\over 8~C_{8L}^{\rm SM}(M_W)}~
{{1\over 3} M_1 (x) + 3 M_2 (x) \over  M_1 (x) } ,
\nonumber  
\\
\chi^{'} & = & 0 .
\end{eqnarray}
Note that $\chi$ and $\chi^{'}$ are functions of $x$ only, whereas $\xi_7 $
depends on $\tilde{m}, x$ and also on $(\delta_{23})_{LR}$.  
Therefore, for a fixed $x$ and assuming $| \delta_{LR} | =  | \lambda_t |$,
one can calculate the $A_{\rm CP}$ as a function of $\tilde{m}$ and $\phi$ 
with the constraints (E1) and (E2).  The result is that only $x \lesssim 1$ 
is consistent with the constraints (E1) and (E2).  As $x$ increases,  
the contribution to $R_{\gamma}$ and/or $R_{g}$ get(s) too large. 

\begin{figure}[ht]
\hspace*{-1.0 truein}
\psfig{figure=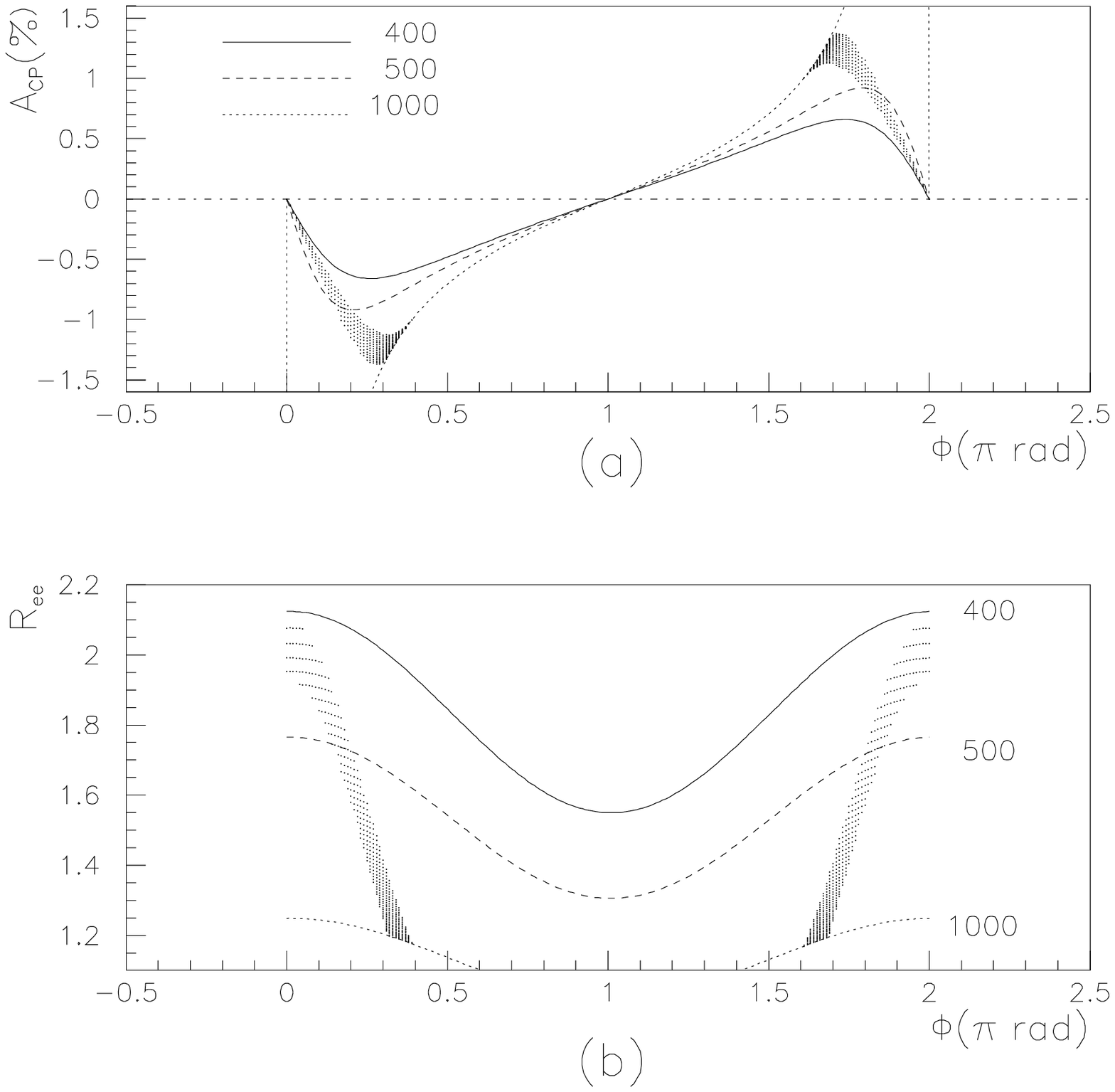}
\caption{ Possible values of $A_{CP}$ and 
$R_{ee}$ as functions of the new phase $\phi$ in the $(LR)$ mixing case
with $x = 0.3$ ({\it i.e.}, $\chi=1.73$ and $\chi^{'} = 0$). 
We assume that $| ( \delta_{23}^{d} )_{LR}| = \lambda_t$.}
\label{fig_lr03}
\end{figure}

\begin{figure}[ht]
\hspace*{-1.0 truein}
\psfig{figure=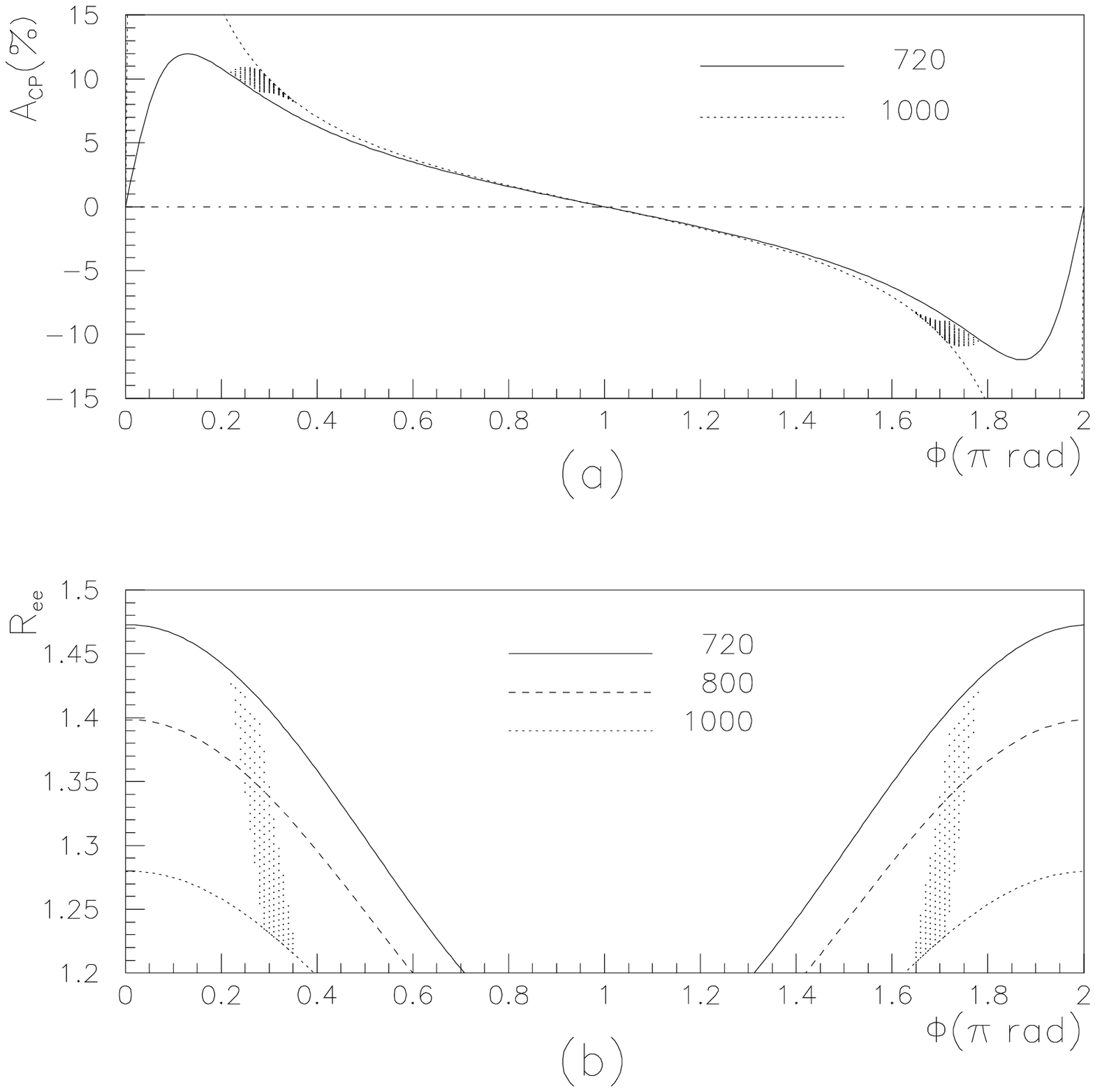}
\caption{The same as Fig.~\ref{fig_lr03} with $x=0.8$ ({\it i.e.}, 
$\chi=5.47$ and $\chi^{'} = 0$).}
\label{fig_lr08}
\end{figure}

\begin{figure}[ht]
\hspace*{-1.0 truein}
\psfig{figure=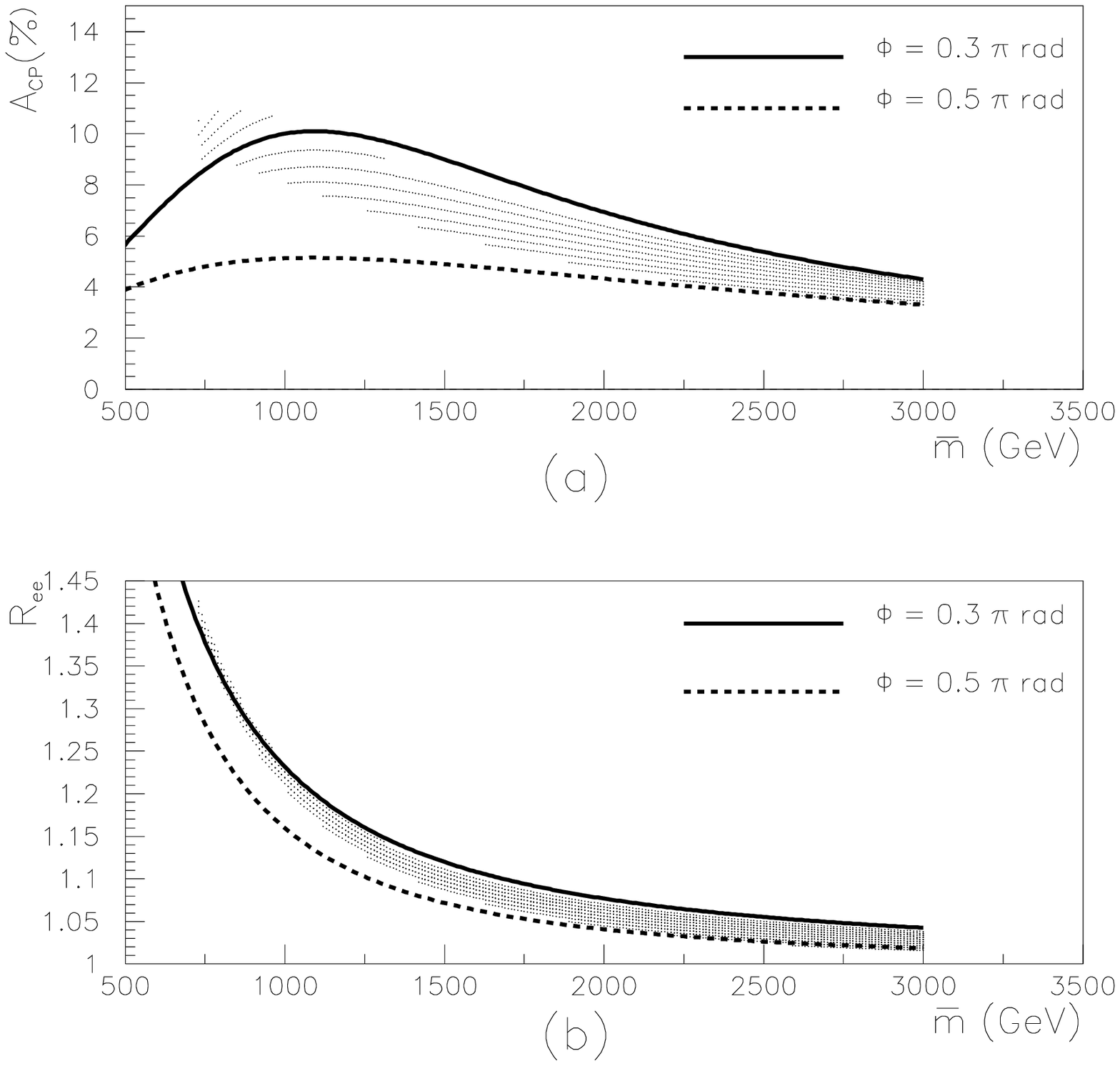}
\caption{(a) $A_{\rm CP}^{b\rightarrow s\gamma}$ and (b) $R_{ee}$ as functions
of the squark mass $\tilde{m}$ in the $(LR)$ mixing case with $x=0.8$.
}
\label{fig_conv}
\end{figure}

For $x=0.3$ and $x=0.8$, the allowed range of $A_{CP}$ and $R_{ee}$ as 
functions of $\phi$ are shown in Fig.~\ref{fig_lr03} and Fig.~\ref{fig_lr08}, 
respectively, along with the constant $\tilde{m}$ contours. 
In the present case where the MIA is appropriate, one also has to take into 
account the constraints on squark masses  from CDF ($\tilde{m} > 230$ GeV 
for $x=1$) \cite{cdf} and D0 ($\tilde{m} > 260$ GeV for $x=1$) \cite{d0}. For 
$x\neq 1$, one can read off the allowed mass range for the squark mass from 
the $(m_{\tilde{g}}, \tilde{m})$ exclusion plot \cite{pdg}.  
Roughly speaking, $\tilde{m} > 300$ GeV for $x=0.3$ and $\tilde{m} > 200$ 
GeV for $x=3.0$. Fig.~\ref{fig_lr03} (a) ($x=0.3$ for which 
$\chi = 1.73, \chi^{'} = 0$) indicates that the direct CP asymmetry 
$A_{\rm CP}^{b\rightarrow s\gamma}$ is in the range $ \lesssim 1.3 \%$ for 
the squark mass  $\tilde{m} = 400-1000$ GeV and the new CP violating phase 
$\phi = 0-0.4\pi$.  This aysmmetry is probably too small to be observed. 
But for the same range of $\tilde{m}$ and $\phi$, the $R_{ee}$ can be as 
large as 2.1 (see Fig.~\ref{fig_lr03} (b)).  So $b\rightarrow s e^+ 
e^-$ is more sensitive to the $(LR)$ mixing than the direct CP asymmetry 
in $b\rightarrow s \gamma$ if $x=0.3$. From Fig.~\ref{fig_lr08} (a) 
($x=0.8$ for which $\chi = 5.47$ and $\chi^{'} = 0$),  the $A_{CP}$ is in 
the range $8-11 \%$ for $| \phi | = 0.2-0.35 \pi$.  It seems that there is 
a definite lower bound to the $A_{CP}$, but this is an artifact due to our
choice of $\tilde{m}< 1$ TeV. For heavier $\tilde{m}$ it vanishes very slowly
(see Fig.~\ref{fig_conv} (a) and the following paragraph). However if all 
the squarks (including the third family squarks) are heavier than $O(1)$ TeV, 
the motivation for the low energy 
SUSY is lost, since the fine tuning problem is reintroduced. Therefore we 
think that the condition $\tilde{m} \lesssim 1$ TeV is a reasonable 
requirement in the scenarios for the soft SUSY breakings where the MIA is
valid. With this caveat, the predicted values for $A_{\rm CP}$ are within 
reach of the B factories.  The impact on $R_{ee}$ is less striking than the 
$x=0.3$ case, but there is still a modest enhancement upto 1.44 of $R_{ee}$ 
over its SM value which may be also detectable at B factories.  

One interesting feature of the $(LR)$ mixing case is that the observables
we show in Figs.~\ref{fig_lr03} and \ref{fig_lr08} can probe the effects of
very heavy squark masses $\tilde{m} = 400 ~(710) - 1000$ GeV for $x = 0.3 ~
(0.8)$. Moreover, the heavier squarks can generate larger $A_{\rm CP}$, 
which may be in conflict with the naive expectation based on the decoupling 
of heavy particles in SUSY models. However, this is just an artifact of 
our requirement $\tilde{m} < 1$ TeV, as described at the end of the previous
paragraph. 
This is because we have fixed $x$, since the heavier squark mass 
$\tilde{m}$ for a fixed $x$ implies the heavier gluino mass $m_{\tilde{g}}$. 
Therefore the $\xi_7$ decreases rather slowly as 
$\tilde{m}$ increases with a fixed $x$ because of the $m_{\tilde{g}}$ factor 
in the numerator of the second term.  In Fig.~\ref{fig_conv} (a) and (b), we 
plot the direct CP asymmetry $A_{\rm CP}$ and $R_{ee}$ as functions of 
$\tilde{m}$ for $x=0.8$. We fixed $\phi = 0.3 \pi$ and $0.5 \pi$. If $\phi$ 
changes its sign, the direct CP asymmetry $A_{\rm CP}$ also changes its sign.
We observe that $A_{\rm CP}$ is maximized around $\tilde{m} = 1$ TeV or so. 
The effects of heavy squarks decouple very slowly for $A_{\rm CP}$ in the 
$(LR)$ mixing case. On the contrary, the effect on $R_{ee}$ is larger for 
the lighter squark mass as usual.   

\subsection{Case (ii) : $(LL)$ insertion }

Next let us consider the case (ii) : $(LL)$ insertion. In this case, the SUSY
breaking terms are the main source of the flavor changing $(LL)$ mixing, which 
are not related with the Yukawa couplings in principle.  Therefore, the MIA
may not be always valid, depending on the superparticle spectra. For example,
a class of models \cite{m1},\cite{m3} falls into this case where the $(LL)$ 
mixing dominates. These models \cite{m1} \cite{m3} predict that 
$(23)$ mixing is order of $\lambda^2$. The mass spectra of the down-squarks 
in the model \cite{m1} are nearly degenerate, whereas in the 
model \cite{m3} only the $\tilde{t}_{L,R}, \tilde{b}_L$, gauginos and the 
lightest neutral Higgs are relatively light compared to $\sim 1$ TeV.
Therefore, one can use the MIA for the first models \cite{m1},
whereas one has to use the vertex mixing for the second model \cite{m3}. 

Below, we will consider the MIA case first. 
In the mass insertion approximation, 
\begin{eqnarray}
\xi_7 & = & 1 + {1\over C_{7L}^{\rm SM} (m_W) }~
{8 \pi \alpha_s \over 9 \sqrt{2} G_F 
\tilde{m}^2 \lambda_t} ( \delta_{23} )_{LL} M_3 (x) ,
\nonumber   \\
\chi & = & {9~C_{7L}^{\rm SM}(M_W)\over 8~C_{8L}^{\rm SM}(M_W)}~
{{1\over 3} M_3 (x) + 3 M_4 (x) \over  M_3 (x) },
\nonumber \\
\chi^{'} & = & { 2 P_1 (x)~C_{7L}^{\rm SM}(M_W) \over M_3 (x)~C_9^{\rm SM}(M_W) }.
\end{eqnarray}
In this case we consider two different choices for $| ( \delta_{23} )_{LL} |$
in order to compare our results with other existing literatures :  
$ | ( \delta_{23} )_{LL} | = | \lambda_t | $ \cite{randall} and 
$| ( \delta_{23} )_{LL} | = O(1)$ \cite{ciuchini}.  
As before, one imposes the experimental informations on $R_{\gamma}$ and 
$R_{g}$, and gets the allowed regions for $A_{\rm CP}$ 
and $R_{ee}$ for a given phase $\phi$, as well as the direct search limit 
on the squark mass from CDF 
and D0. 
For $x=0.3,1.0$ and 3.0, the $(\chi, \chi^{'}) = (7.28,0.19), (5.25, 
0.27)$ and $(3.83, 0.40)$, respectively. Therefore, the overall features of 
various observables will be close to Fig.~\ref{fig_ch5pr03}, except that 
$\xi_7$ should be fixed to some definite value.   

In case $| ( \delta_{23} )_{LL} | = | \lambda_t |$, there are no visible 
new physics effects on $A_{\rm CP}$ and $R_{ee}$, and 
the CP violating dilepton signals can be complimentary to our study.
One always have 
\begin{equation}
| A_{\rm CP} | \lesssim 1 \%,~~~ | R_{ee} - 1 | \lesssim 0.01~~~({\rm for}~
| ( \delta_{23} )_{LL} | = | \lambda_t | )
\end{equation} 
after all the experimental constraints are imposed. 
However there may be some visible deviation in 
$A_{\rm FB} (ee)$ as inferred from Fig.~\ref{fig_ch5pr03}.           

\begin{figure}[ht]
\hspace*{-1.0 truein}
\psfig{figure=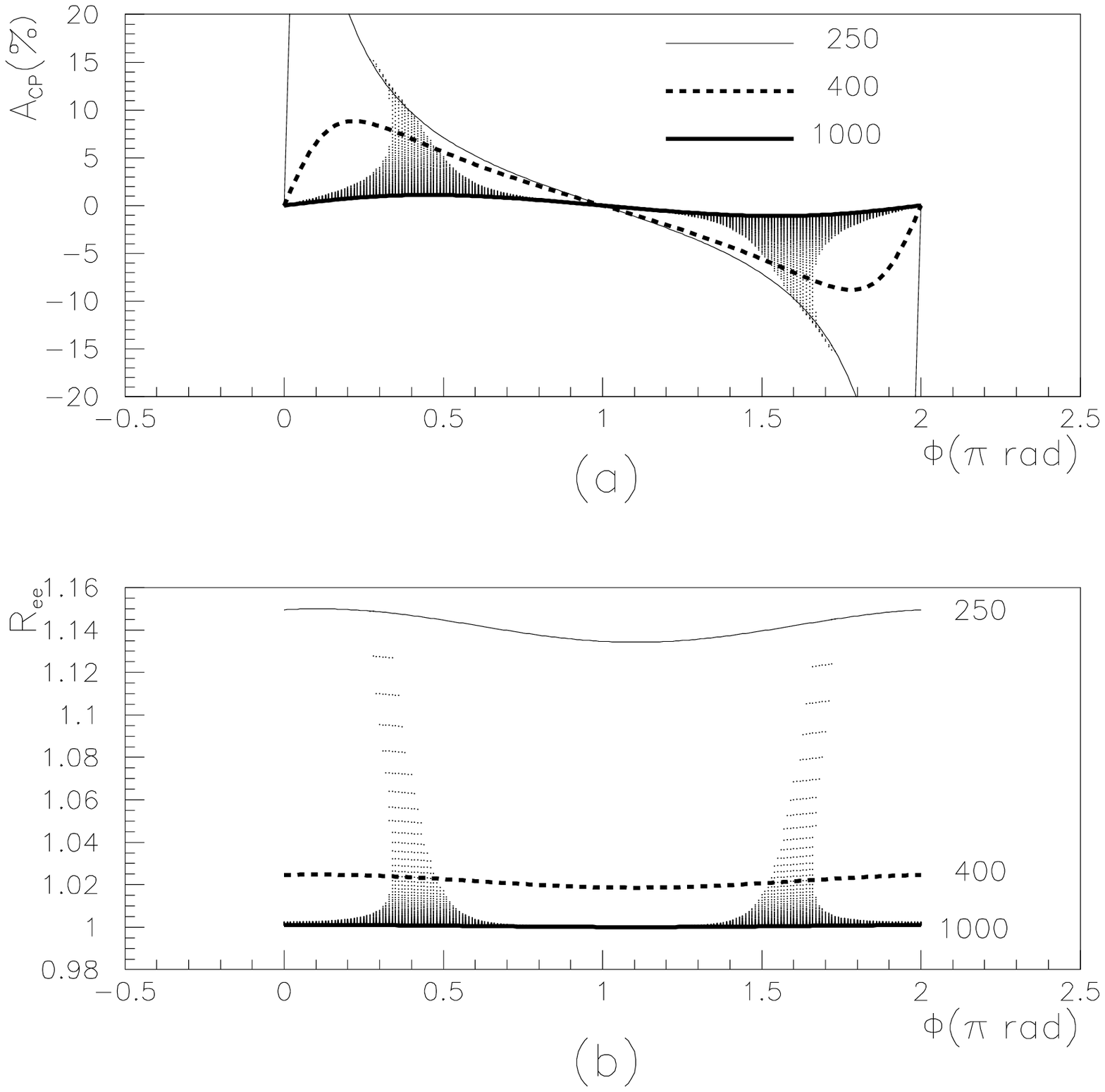}
\caption{ Possible values of $A_{CP}$ and 
$R_{ee}$ as functions of the new phase $\phi$ in the $(LL)$ mixing case
with $x = 0.3$ ({\it i.e.}, $\chi=7.28$ and $\chi^{'} = 0.19$). 
We assume that $| ( \delta_{23}^{d} )_{LL}| = 1$.}
\label{fig_px03}
\end{figure}

\begin{figure}[ht]
\hspace*{-1.0 truein}
\psfig{figure=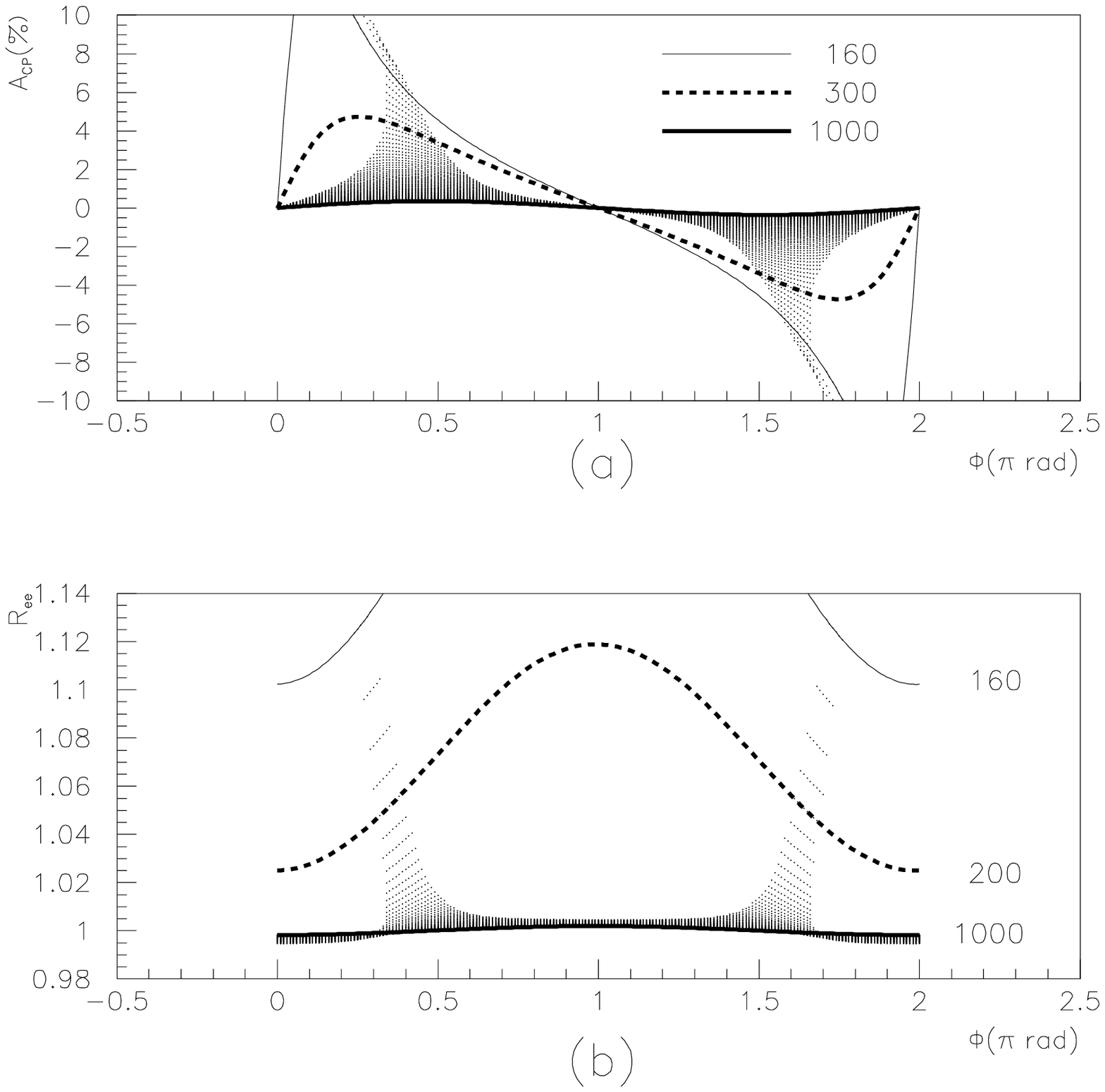}
\caption{ Possible values of $A_{CP}$ and 
$R_{ee}$ as functions of the new phase $\phi$ in the $(LL)$ mixing case
with $x = 1$ ({\it i.e.}, $\chi=5.25$ and $\chi^{'} = 0.27$). 
We assume that $| ( \delta_{23}^{d} )_{LL}| = 1$.}
\label{fig_px1}
\end{figure}

If  $| ( \delta_{23} )_{LL} | \sim O(1)$ as assumed in Ref.~\cite{ciuchini}, 
then one expects that $A_{CP}$  can be as large as  $\pm 10 \%$ to 
$\pm 15 \%$ for $\phi \sim \pm 0.3 \pi$ for $x \sim 1 - 0.3$, although
the $R_{ee}$ does not change very much from its SM value (Figs.~
\ref{fig_px03} and \ref{fig_px1}).  If $x$ gets larger, the $A_{\rm CP}$ 
gets smaller and eventually becomes undetectable ({\it e.g.}, 
$A_{\rm CP} \lesssim 2 \%$ for $x = 3$, if we impose $\tilde{m} > 200$ GeV). 
In Ref.~\cite{ciuchini}, it was noted that this new CP violating phase 
could result in the CP violation in the decay amplitudes for 
$B\rightarrow (\phi, \pi^0) 
+ K_S$ at the level of $(0.1-0.7)$ of the SM amplitude depending on the 
squark mass. There would be some intrinsic theoretical uncertainties in such
estimates of nonleptonic exclusive $B$ decays. On the contrary, the direct 
CP violation in $b\rightarrow s\gamma$ can provide independent informations 
on $| ( \delta_{23} )_{LL} |$ with less theoretical uncertainties, since we 
are dealing with the inclusive decay rate. In any rate the observable 
$A_{\rm CP}$ can play an important role in probing a new CP violating phase 
in $B$ decays if  the condition $| ( \delta_{23} )_{LL} | \sim O(1)$ is met.  

The vertex mixing case can be obtained from Eq.~(40) by following 
identifications :
\begin{eqnarray}
( \Gamma_{GL}^d )_{I,j} & = & \left\{ 
\begin{array}{cc}  
( V_{L} )_{i,j} & ({\rm for} ~ I=1,2,3)  \\ 0 & ({\rm otherwise}),
\end{array} 
\right.
\nonumber \\
( \Gamma_{GR}^d )_{I,j} & = & \left\{ 
\begin{array}{cc} 
0 & ({\rm for}~ I = 1,2,3) \\
 - ( V_{R} )_{i,j} & ({\rm otherwise}).  \\
\end{array}
\right.
\end{eqnarray}
In the $(LL)$ mixing case we consider here, one has $(V_R)_{ij}=\delta_{ij}$. 
Also, we assume that $| ( V_{L} )_{23} | = O(0.1)$ with a new phases of 
$O(1)$.  This assumption is motivated by a recent model by Kaplan 
{\it et al.} \cite{u1}, which is  a SUSY model of flavor based on the single 
$U(1)$ generating the fermion spectra as well as communicating SUSY breaking 
to the visible sector.  In this model, only the third generation squarks are
lighter than $\sim 1$ TeV, and the 1st and the 2nd generation squarks simply
decouple. Therefore,  we can keep only the third family squarks ($I=3$)
in the sum over the squark mass eigenstates, since others are all heavier 
than $O(1)$ TeV and/or the relations (40),(49) hold.  After one imposes the 
experimental informations on $R_{\gamma}$ and $R_{g}$, one gets the allowed
regions for $A_{CP}$ and $R_{ee}$ for a given phase $\phi$, 
as shown in Figs. 10 (a)-(c) for $x=0.3, 1.0$ and 3.0. 
The corresponding values of $(\chi,\chi^{'})$ are $(9.27,0.16), 
(7.50, 0.20)$ and $(6.20, 0.26)$, respectively. In Fig.~10, 
we superposed the contours for three different values of $\tilde{m}_3=100, 
200, 1000$ GeV. In the case only the third generation squarks are light, 
the strongest bound on the lighter stop comes from LEP experiments 
\cite{boer}, and $\tilde{m}_3=100$ GeV is not excluded yet by  
LEP experiment. Therefore, one expects that $A_{\rm CP}$  
can be as large as 6 $\%$ to 12 $\%$ for $\phi \sim 0.7 \pi$ radian for 
$x=0.3, 1.0$ and 3.0 respectively, although the $R_{ee}$ does change very 
little : $1 \lesssim R_{ee} \lesssim 1.1$.  However, this large $A_{\rm CP}$
quickly diminishes as $\tilde{m}_{3}$ gets heavier, and $| A_{\rm CP} | 
\lesssim 2 \%$ for $\tilde{m}_{3} > \sim 200$ GeV. 

Therefore, it is very difficult to see the effects of the $(LL)$ insertion 
in the effective SUSY models (for which the VM method is valid) as well as 
the case of almost degenarate squarks (for which MIA is  valid) from 
$R_{ee}$.  In other words, the $(LL)$ insertion can generate a large direct
CP violation in $b \rightarrow s \gamma$ if there is a new CP violating phase
associated with the squark mass matrix, $( M_{LL}^{\tilde{d}} )^2$, whereas
there can be no significant change in $R_{ee}$ compared with the SM case.
Also the deviation from the SM diminish very quickly as stop gets heavier.
Practically speaking, it would be impossible to  notice the new physics 
effects if $\tilde{m} > 200$ GeV unless $|\delta_{LL}| \sim O(1)$, for which 
new physics signal can be visible for heavier stop until $\tilde{m} \sim 400$
GeV if $x$ is not too large. 
This is in contrast to the $(LR)$ mixing case, for which the new physics 
effects can increase $R_{ee}$ up to 2.15, and the $A_{\rm CP}$ can be as 
large as 11 $\%$ for fairly large stop, $\tilde{m} \sim 1000 $ GeV and 
$x = 0.8$ (see Fig.~6).  
Although $A_{\rm CP}$ and $R_{ee}$ are not sensitive to 
the $(LL)$ insertion for $| \delta_{LL} | \sim \lambda_t$, there is another
observable which is complementary to our observables : namely, CP violating
lepton asymmetry in B decays discussed in Ref.~\cite{randall}. For larger
$ | \delta_{LL} | \sim O(1)$, direct CP violations in nonleptonic $B$
decays through $\Delta B = 1$ penguin operators can provide additional 
informations \cite{ciuchini}.
Again, different channels are sensitive to different types of new physics,
and it will be helpful to study as many modes as possible in order to find
out new physics signals at B factories.  


\section{Conclusions}

In conclusion, we considered the possible new physics effects on the 
$b\rightarrow s l^+ l^-$ through the modified $b\rightarrow s \gamma$ vertex. 
The CP violation in $b \rightarrow s \gamma$ can be very different from the 
SM expectation ($A_{\rm CP} ({\rm SM}) \simeq 0$), 
and the branching ratio and $A_{\rm FB}$ in $b\rightarrow s l^+ l^-$ 
can be affected by the new physics contributing to $b\rightarrow s \gamma$.  
In particular, the usual model-independent extraction of the Wilson 
coefficients $C_{7,9,10}$ may be useless in the presence of new physics 
that modifies the $C_{7,8}$ with a new CP-violating phase (namely, 
Im $(\xi_7) \neq 0$). 
Therefore, not only is the CP asymmetry in $b \rightarrow s \gamma$  a 
sensitive probe of new physics that might be discovered 
at $B-$factories, but also it is indispensable for the model-independent 
analysis of $b\rightarrow s l^+ l^-$. Search for $A_{\rm CP}$  is clearly 
warranted at $B-$factories. 

We also considered specific models which satisfy our assumptions made in the 
model-independent analysis : namely, generalized SUSY models with 
gluino-mediated FCNC. In the case of $(LR)$ mixing, $R_{ee}$ can be enhanced
compared to the SM value. Also the direct asymmetry $A_{\rm CP}$ can be as 
large as $8-11 \%$ for $x = 0.8$ and $\tilde{m} = 1$ TeV. 
In this case, the direct asymmetry $A_{\rm CP}$ is sensitive to the heavy 
squark masses, since the decoupling occurs very slowly, beyond $\tilde{m} =
1.1 $ TeV (see Fig.~7 (a)). Also there is a lower bound on  $A_{\rm CP}$ 
since all the squarks cannot be simultaneously heavier than $O(1)$ TeV.
This is quite an intersting feature of the $(LR)$ mixing scenario.
In the $(LL)$ mixing case, there is no observable effects both for 
$A_{\rm CP}$  and $R_{ee}$ if $| \delta_{LL} | = | \lambda_t |$. 
But there can be an appreciable amount of $A_{\rm CP}$ upto $\pm 15 \%$, 
if $| \delta_{LL} | = O(1)$ in the MIA.  In the $(LL)$ mixing with the VM 
approximation, one may be observable $A_{\rm CP}$ upto $6-12 \%$ depending on 
$x$ and the new CP violating phase $\phi$ for $| ( V_{L} )_{23} | \sim 0.1$
which is the typical values in the model by Kaplan {\it et al.} \cite{u1}.

Our study is also complimentary to other previous works, {\it e.g.}, 
the dilepton asymmetry considered by Randall and Su \cite{randall}, and 
the CP asymmetry in the decay amplitudes for nonleptonic $B$ decays
considered by Ciuchini {\it et al.} \cite{ciuchini}.  It is very  important 
to measure various kinds of CP asymmetries at B factories, 
especially those CP  asymmetries which (almost) vanish in the SM like the 
direct asymmetry in $b\rightarrow s \gamma$ and dilepton asymmetry, 
in order to probe new CP violating phase(s) that may be necessary for us 
to understand the baryon number asymmetry of the universe.  
Different channels may be sensitive to different parameter values in 
new physics, and thus can provide indepedent informations on new physics. 

While we were preparing this manuscript, we received a preprint by 
Chua {\it et al.} \cite{chua} considering  the CP-violation in 
$b\rightarrow s \gamma$ in supersymmetric models.  It somewhat overlaps 
with Sec. IV of our present work. But they did not consider the 
$b\rightarrow s g$ constraint, and get somewhat larger $A_{CP}$ asymmetry 
than our work. 

\acknowledgements

Y.K. and P.K. acknowldge the hospitality of Korea Institute for Advanced 
Study (KIAS) where a part of this work has been done.  
This work is supported in part by KOSEF Contract No. 971-0201-002-2, 
by KOSEF through Center for Theoretical Physics at Seoul National University, 
by the Ministry of Education through the Basic Science Research Institute,
Contract No. BSRI-98-2418, the German-Korean scientific exchange 
programme DFG-446-KOR-113/72/0 (PK), and the KAIST Center for Theoretical 
Physics and Chemistry (PK,YK).


%
%

\begin{figure}[ht]
\hspace*{-1.0 truein}
\psfig{figure=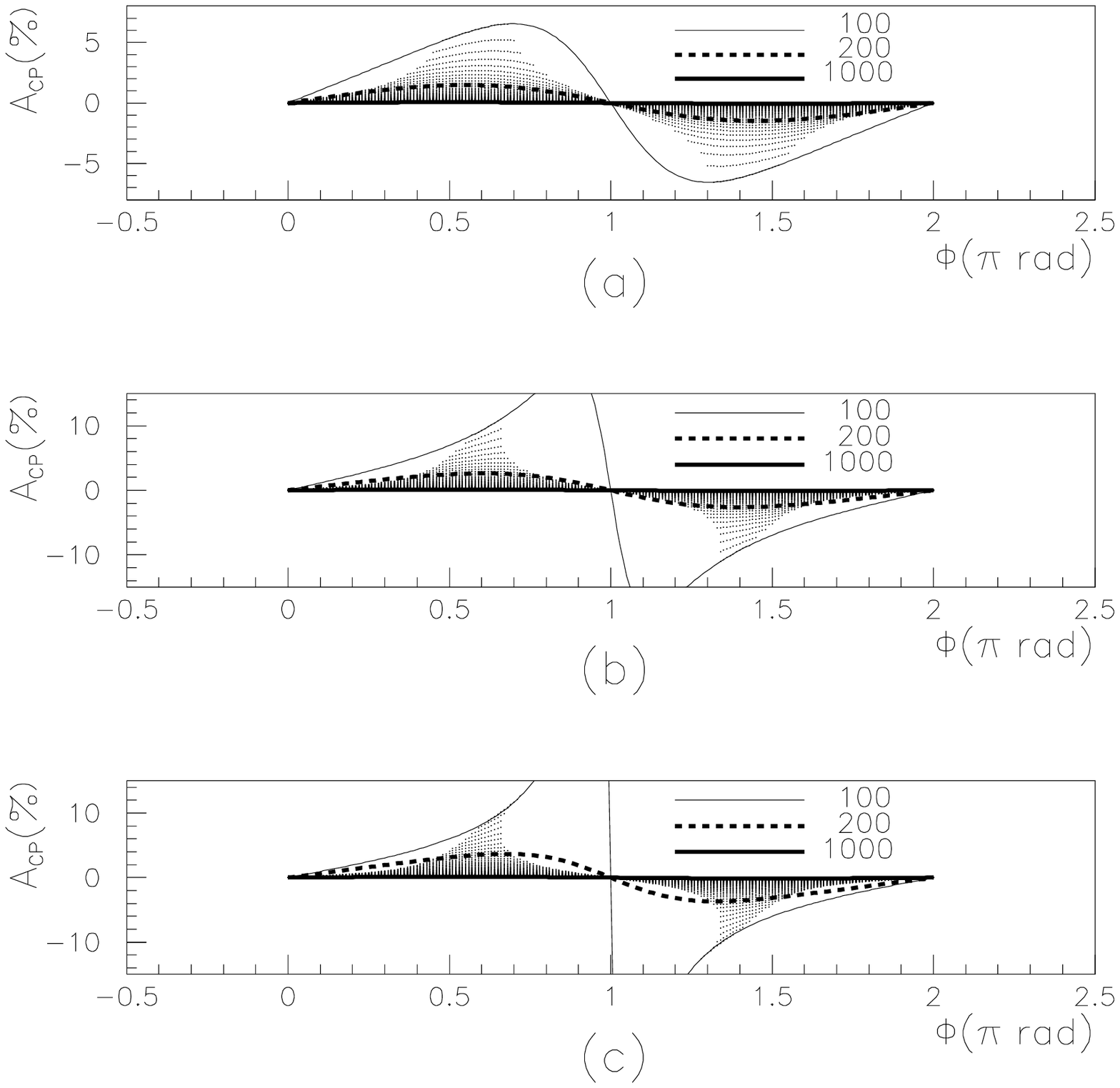}
\caption{The $A_{\rm CP}$ contours in the $(\tilde{m},
\phi)$ plane for (a) $x = 0.3$, (b) $x=1$ and (c) $x=3$  in the $(LL)$ 
insertion case using the vertex mixing method. }
\label{fig_vx}
\end{figure}


%

\end{document}